\documentclass[a4paper,11pt]{article}
\usepackage[left=2cm,right=2cm,top=2cm,bottom=2cm]{geometry}
\usepackage[usenames,dvipsnames,svgnames,table]{xcolor}
\usepackage{psfrag,amsmath,amsfonts,amssymb,latexsym,amsthm,verbatim,color,lscape,multirow,graphicx,subcaption}
\usepackage[figuresright]{rotating}
\usepackage[compress]{natbib}
\usepackage{url}
\usepackage{hyperref}
\hypersetup{colorlinks,
citecolor=black,
filecolor=black,
linkcolor=black,
urlcolor=black,
pdftex}
\usepackage{tikzducks}
\usepackage{booktabs}
\usepackage[displaymath, mathlines,pagewise]{lineno}
\usepackage{colortbl}
\definecolor{gray}{rgb}{0.9,0.9,0.9}

\let\proglang=\textsf
\newcommand{\pkg}[1]{{\fontseries{b}\selectfont #1}}

\usepackage{caption}

\usepackage{times}
\usepackage{blindtext}
\usepackage[font={footnotesize,it}]{caption}

\usepackage{sectsty}  
\subsectionfont{\normalsize\bf}
\sectionfont{\large\bf}

\def\C{\mathbf{C}}

\def\pp{\mathbf{p}}

\def\y{\mathbf{y}}

\def\bfa{\mathbf{a}}
\def\X{\mathbf{X}}

\def\de{\delta}
\def\debf{\boldsymbol{\delta}}

\def\pibf{\boldsymbol{\pi}}

\def\th{\theta}
\def\thbf{\boldsymbol{\theta}}
\def\pibf{\boldsymbol{\pi}}
\def\Xibf{\boldsymbol{\Xi}}

\def\C{\mathbf{C}}

\def\pp{\mathbf{p}}

\def\y{\mathbf{y}}

\def\de{\delta}
\def\Debf{\boldsymbol{\Delta}}

\def\th{\theta}
\def\thbf{\boldsymbol{\theta}}

\def\pibf{\boldsymbol{\pi}}
\def\Xibf{\boldsymbol{\Xi}}

\long\def\symbolfootnote[#1]#2{\begingroup
\def\thefootnote{\fnsymbol{footnote}}\footnote[#1]{#2}\endgroup}

\usepackage[normalem]{ulem}

\begin{document}

\title{Bi-factor and second-order copula models  
for item response data}

\author{Sayed H. Kadhem  \and Aristidis~K.~Nikoloulopoulos \footnote{Correspondence to: \href{mailto:a.nikoloulopoulos@uea.ac.uk}
{a.nikoloulopoulos@uea.ac.uk}, Aristidis K. Nikoloulopoulos, School of Computing Sciences, University of East Anglia, Norwich NR4
7TJ, U.K.} }

\date{}

\maketitle

\begin{abstract}
\baselineskip=24pt
 \noindent
 Bi-factor and second-order models based on copulas are proposed for item response data,  where the items can be split into non-overlapping groups such that there is a homogeneous dependence within each group. Our general models include the Gaussian bi-factor and second-order models as  special cases and can lead to more probability in the joint upper or lower tail compared with the Gaussian bi-factor and second-order models.  Details on maximum likelihood estimation of parameters for the bi-factor  and second-order copula models are given, as well as model selection and goodness-of-fit techniques.
Our general methodology is demonstrated with an extensive simulation study and illustrated for the Toronto Alexithymia Scale. Our studies suggest that there can be a substantial improvement over the Gaussian bi-factor and second-order models both conceptually, as the  items can have interpretations of latent maxima/minima or mixtures of means in comparison with latent means, and in fit to data.   


\end{abstract}

\baselineskip=24pt

\section{Introduction}
\noindent Factor models are statistical frameworks that analyse multivariate observed variables using few latent variables or factors \citep{Bartholomew&Knott&Moustaki2011-Wiley}. They are increasingly popular in psychometrics applications, where constructs  such as general well-being could be assessed via survey questions (referred to as items).  Datasets with large number of items are naturally divided into subgroups, in such, each group of items has homogeneous dependence. For example, the well-being (common factor) of patients is usually assessed via items that arise from several sub-domains to assess several group-specific factors such as the depression, anxiety and stress.  
This special classification of items is also common in educational assessments and termed as testlets \citep{Wainer&Kiely1987}. It is essential to investigate the items structure, as implementing factor models on testlet-based items could result in biased estimates and poor fit \citep{Wang&Wilson2005,DeMars2006,Zenisky-etal2002,Sireci-etal1991,Lee&Frisbie1999,Wainer&Thissen1996}.

To account for the homogeneous dependence in each group of items,
\cite{Gibbons1992-PKA} and \cite{Gibbons-etal2007} proposed bi-factor models for binary and ordinal response data, respectively. The bi-factor models have become omnipresent in analysing  survey items that arise from several sub-domains or groups. 
They consist of a common factor that is linked to all items, and non-overlapping group-specific factors. The common factor explains dependence between items for all groups, while the group-specific factors explain dependence amongst items within each group. The items  are assumed to  be independent given the group-specific and common factors. 

An alternative way of modelling items that are split into several groups is via 
the second-order  model (e.g., \citealt{
delaTorre&Song09,Rijmen2010}), where items are indirectly mapped to 
an overall (second-order) factor  via non-overlapping group-specific (first-order) factors. 
Second-order models are suitable when  the first-order factors are associated  with each other, and there is a second-order factor that accounts for the relations among the first-order factors.

The bi-factor and the second-order models are not generally equivalent   \citep{Yung-etal1999,Gustafsson&Balke1993,
Mulaik&Quartetti1997,Rijmen2010}, unless proportionality constraints are imposed by using the Schmid-Leiman transformation method \citep{Schmid&Leiman1957}. 
More importantly, both models are restricted to the MVN assumption for the latent variables, which might  not  be valid. 
\cite{Nikoloulopoulos2015-PKA}  emphasized that if the ordinal variables in item response can be thought of as discretization of latent random variables that are maxima/minima or mixtures of means 
then the use of factor models based on the MVN assumption for the latent variables could provide poor fit. 

In the context of item response data, latent maxima, minima and means can arise
depending on how a respondent considers specific items.
An item might make the respondent think about $M$ past events which, say, have
values $W_1,\ldots,W_M$. In answering the item, the subject
might take the average, maximum or minimum  of $W_1,\ldots,W_M$
and then convert to the ordinal scale depending on the magnitude.
The case of a  latent maxima/minima can occur if the response
is based on a best or worst case. 
For different dependent items based on latent maxima or minima, multivariate extreme value and copula theory can be used to select suitable distributions for the latent variables. Copulas that arise from extreme value theory have more probability in one joint tail (upper or lower) than expected with a   MVN distribution where and have latent variables that are maxima/minima instead of means.
Even, in the case where  the item responses are based on discretizations of latent variables that are
means, then it is possible that there can be more probability in
both the joint upper and joint lower tail, compared with MVN distributed latent variables. 
This happens if the respondents consist of a ``mixture" population
(e.g., different locations or genders). From the theory of
elliptical distributions and copulas, it is known that the multivariate Student-$t$ distribution as a scale mixture
of MVN has more dependence in the tails.

 \cite{Nikoloulopoulos2015-PKA} have studied factor copula  models for item response  and have shown that there is an improvement on the factor models based on the MVN assumption for the latent variables both conceptually and in fit to data. This improvement relies on the aforementioned  reasons, i.e.,  items  can have more probability in
joint upper
or lower tail than would be expected with a  MVN or 
items can be considered as discretized maxima/minima or mixtures of discretized means
rather than discretized means. 

In this paper, we propose copula extensions for bi-factor and second-order models. 
The construction of the bi-factor copula model 
exploits the use of  bivariate copulas that link the observed variables to the common and group-specific factors. Note that if there is only one group of items,  then the bi-factor model reduces to the 2-factor copula model in \cite{Nikoloulopoulos2015-PKA}.
Similarly with the bi-factor copula model, we also use bivariate copulas to construct the second-order copula model. In this case, there are  bivariate  copulas that link the observed to the group-specific factors, and also bivariate  copulas that link the group-specific to the second-order factor. To account for the dependence between the observed variables and group-specific factors, each group of variables in fact is modelled using the one-factor copula model proposed by \cite{Nikoloulopoulos2015-PKA}. In addition, 
 if there is only one group of items, then the second-order copula model reduces to the one-factor copula model. Hence, the proposed models contain the one- and two-factor copula models in \cite{Nikoloulopoulos2015-PKA} as special cases, while allowing flexible dependence structure for both within  and between group dependence. As a result, the models are suitable for modelling a high-dimensional  item response  classified into non-overlapping groups.

The proposed models  copula models 
are truncated 
vine copulas \citep{Brechmann-Czado-Aas-2012} that involve both observed and latent variables.   
They provide flexible dependence by selecting arbitrary bivariate linking copulas \citep{Joe&Li&Nikoloulopoulos2010} to link the items to latent factors. If the bivariate linking copulas are BVN, then the Gaussian bi-factor  
and   second-order models 
are special cases of our constructions which are the discrete counterparts of the structured factor copula models introduced by \cite{Krupskii&Joe-2015-JMVA}.

The remainder of the paper proceeds as follows. Section \ref{sec-model}  introduces
the  bi-factor and second-order  copula models for item response and discusses their relationship with the existing models. Estimation techniques and computational details are provided in Section \ref{sec-computing}. Section \ref{sec-selection}  proposes simple diagnostics based on semi-correlations and an heuristic method to  select suitable bivariate copulas and build  plausible bi-factor and second-order copula models. 
Section \ref{sec-gof} summarizes the assessment of goodness-of-fit of 
these models  using
the $M_2$ statistic of \cite{Maydeu-Olivares&Joe2006}, which is based on a quadratic
form of the deviations of sample and model-based proportions over
all bivariate margins.
Section \ref{sec-sim}  contains an extensive simulation study to gauge the small-sample efficiency of the proposed estimation, investigate the misspecification of the bivariate copulas, and examine the reliability of the model selection and goodness-of-fit techniques.
Section \ref{sec-app} presents an  application
of our methodology to the Toronto Alexithymia Scale. 
In this example, it turns out that our models, with linking copulas
selected according to the items being  latent minima or
mixtures of means, provide better fit than the Gaussian bi-factor and second-order models. 
We conclude with some discussion in Section \ref{sec-disc}, followed by a technical Appendix.

\baselineskip=27pt

\section{\label{sec-model}Bi-factor and second-order copula models}
 
Let $\underbrace{Y_{11}, \ldots,Y_{d_11}}_1,\ldots, \underbrace{Y_{1g},\ldots, Y_{d_gg}}_g, \ldots, \underbrace{Y_{1G}, \ldots, Y_{d_GG}}_G$ denote the item response variables classified into the $G$ non-overlapping groups.  There are $d_g$ items in group $g$; $g=1,\ldots,G,\,j=1,\ldots,d_g$ and collectively there are $d=\sum_{g=1}^{G} d_g$ items, which are all measured on an ordinal scale; $Y_{jg}\in \{0,\ldots,K_{jg}-1\}$. 
Let the cutpoints in the uniform $U(0,1)$ scale for the $jg$'th item be $a_{jg,k}$, $k=1,\ldots,K-1$,  with $a_{jg,0}=0$ and $a_{jg,K}=1$. These correspond to $a_{jg,k}=\Phi(\alpha_{jg,k})$, where $\alpha_{jg,k}$ are cutpoints in the normal $N(0, 1)$ scale.

The bi-factor and second-order factor copula models are presented in Subsections \ref{bifactor-subsec} and \ref{2ndorder-subsec}, respectively. 
Subsection \ref{special-subsec} discusses their relationship with the existing Gaussian bi-factor and second-order models.

\subsection{\label{bifactor-subsec}Bi-factor copula model}
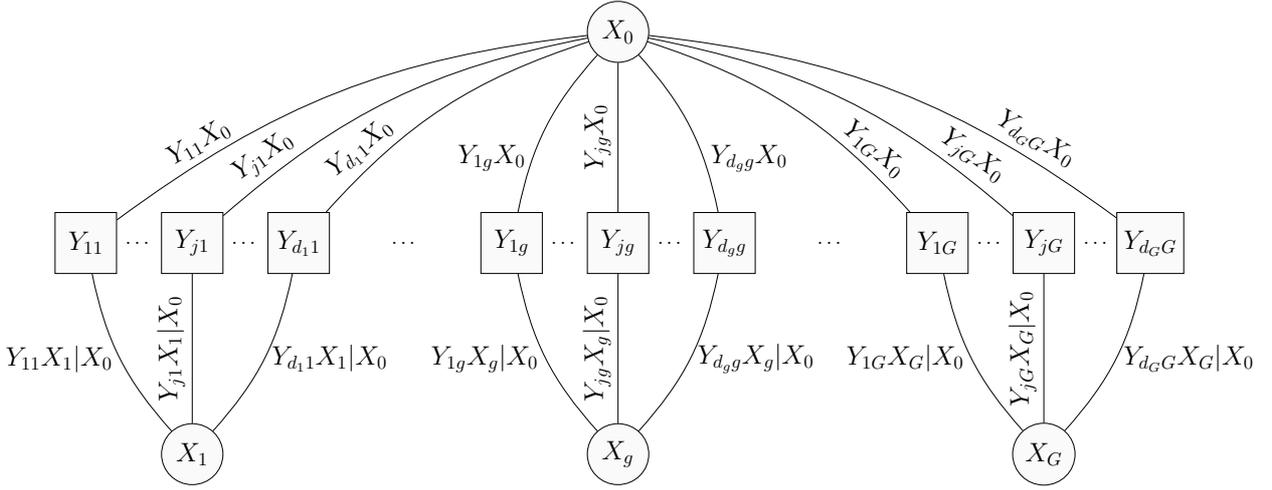
\begin{figure}[h!]
\scalebox{.7}{
    \begin{tikzpicture}
        [square/.style={
            draw,
            fill=gray!20,
            minimum width=3em,
            minimum height=3em,
            node contents={#1}}
            ]

        \node at (0,1) [circle,draw,fill=gray!20,minimum width=3em] (X0) {\Large $X_0$};
       
    		\node at (-8,-7) [circle,draw,fill=gray!20,minimum width=3em] (X1) {\Large $X_1$};
    		        
    		\node at (0,-7) [circle,draw,fill=gray!20,minimum width=3em] (Xg) {\Large $X_g$};

    		\node at (8,-7) [circle,draw,fill=gray!20,minimum width=3em] (XG) {\Large $X_G$};

	    \node at (-10,-3) (Y11) [square={\Large $Y_{11}$}];
         		\node at  (-9,-3)  {$\cdots$};
        \node at (-8,-3) (Yj1) [square={\Large $Y_{j1}$}];
                  \node at  (-7,-3)  {$\cdots$};
        \node at (-6,-3) (Yd11) [square={\Large $Y_{d_11}$}];
	    
\path(X0)edge[black,bend right=15] node[above,rotate=30, midway,pos=0.80]{\Large $Y_{11}X_0$}  (Y11);
\path(X0)edge[black,bend right=15] node[above,rotate=35, midway,pos=0.85]{\Large $Y_{j1}X_0$} (Yj1);
\path(X0)edge[black,bend right=15] node[above,rotate=40, midway,pos=0.80]{\Large $Y_{d_11}X_0$} (Yd11);
\node at  (-4,-3)  {$\cdots$};
	
        \node at (-2,-3) (Y1g) [square={\Large $Y_{1g}$}];
         		\node at  (-1,-3)  {$\cdots$};
        \node at (0,-3) (Yjg) [square={\Large $Y_{jg}$}];
                  \node at  (1,-3)  {$\cdots$};
        \node at (2,-3) (Ydgg) [square={\Large $Y_{d_gg}$}];
	    
\path(X0)edge[black,bend right=15] node[left,rotate=0, midway,pos=0.7]{\Large $Y_{1g}X_0$} (Y1g);
\path(X0)edge[black,bend right=0] node[above,rotate=90, midway,pos=0.50]{\Large $Y_{jg}X_0$} (Yjg);
\path(X0)edge[black,bend right=-15] node[right,rotate=0, midway,pos=0.7]{\Large $Y_{d_gg}X_0$} (Ydgg);
\node at  (4,-3)  {$\cdots$};
	
        \node at (6,-3) (Y1G) [square={\Large $Y_{1G}$}];
         		\node at  (7,-3)  {$\cdots$};
        \node at (8,-3) (YjG) [square={\Large $Y_{jG}$}];
                  \node at  (9,-3)  {$\cdots$};
        \node at (10,-3) (YdGG) [square={\Large $Y_{d_GG}$}];
	    
\path(X0) edge[black,bend right=-15] node[above,rotate=-40, midway,pos=0.8]{\Large $Y_{1G}X_0$} (Y1G);
\path(X0) edge[black,bend right=-15] node[above,rotate=-35, midway,pos=0.85]{\Large $Y_{jG}X_0$} (YjG);
\path(X0) edge[black,bend right=-15] node[above,rotate=-30, midway,pos=0.8]{\Large $Y_{d_GG}X_0$} (YdGG);	    
        \path(X1) edge[black,bend left=15] node[left,midway,pos=0.5]{\Large $Y_{11}X_1|X_0$} (Y11);	    
        \path(X1) edge[black] node[above,midway,rotate=90,pos=0.5]{\Large $Y_{j1}X_1|X_0$} (Yj1);	    
        \path(X1) edge[black,bend right=15] node[right,midway,pos=0.5]{\Large $Y_{d_11}X_1|X_0$} (Yd11);	    
        \path(Xg) edge[black,bend left=15] node[left,midway,pos=0.5]{\Large $Y_{1g}X_g|X_0$} (Y1g);	    
        \path(Xg) edge[black] node[above,midway,rotate=90,pos=0.5]{\Large $Y_{jg}X_g|X_0$} (Yjg);	   
        \path(Xg) edge[black,bend right=15] node[right,midway,pos=0.5]{\Large $Y_{d_gg}X_g|X_0$} (Ydgg);	    
        \path(XG) edge[black,bend left=15] node[left,midway,pos=0.5]{\Large $Y_{1G}X_G|X_0$} (Y1G);	    
        \path(XG) edge[black] node[above,midway,rotate=90,pos=0.5]{\Large $Y_{jG}X_G|X_0$} (YjG);	   
        \path(XG) edge[black,bend right=15] node[right,midway,pos=0.5]{\Large $Y_{d_GG}X_G|X_0$} (YdGG); 	  

    \end{tikzpicture}
}
\caption{\label{bifactor-graph}Graphical representation of the bi-factor copula model with $G$ group-specific factors and a common factor $X_0$. }
\end{figure}

Consider a common factor $X_0$ and $G$ group-specific factors $X_1,\ldots,X_G$, where $X_0,X_1,\ldots,X_G$ are independent and standard uniformly distributed. Let $Y_{jg}$ be the $j$th observed variable in group $g$, with $y_{jg}$ being the realization. 
The bi-factor model assumes that $Y_{1g},\ldots, Y_{d_gg}$ are conditionally independent given $X_0$ and $X_g$, and that  $Y_{jg}$ in group $g$ does not depend on $X_{g'}$ for $g\neq g'$. Figure \ref{bifactor-graph} depicts a graphical representation of the model.

The joint probability mass function (pmf) is given by
\begin{eqnarray*}
\pi(\y)&=&\Pr(Y_{jg} = y_{jg} ; j=1,\ldots,d_g,g=1,\ldots,G)  \\
&=&\int_{[0,1]^{G+1}} \prod_{g=1}^{G}\prod_{j=1}^{d_g} \Pr(Y_{jg} = y_{jg}| X_0 =x_0, X_g=x_g) d{x_1}\cdots d{x_G} d{x_0}.
\end{eqnarray*}
According to Sklar's theorem \citep{Sklar1959} there exists a bivariate copula $C_{Y_{jg},X_0}$ such that $\Pr(Y_{jg} \leq y_{jg}, X_0 \leq x_0)= C_{Y_{jg},X_0}\bigl(F_{{Y_{jg}}}(y_{jg}), x_0\bigr)$,
 for $x_0 \in [0,1]$, where $C_{Y_{jg},X_0}$ is the copula that links observed variable with the common factor $X_0$, $F_{Y_{jg}}$ is the cumulative distribution function (cdf) of  $Y_{jg}$; note that $F_{Y_{jg}}$ 
is a step function with jumps at
$0,\ldots,K-1$, i.e.,   $F_{Y_{jg}}(y_{jg})=a_{jg,y_{jg}+1}$. Then it follows that, 
$$
F_{Y_{jg}|X_0}(y_{jg}|x_0):= \Pr(Y_{jg} \leq y_{jg} | X_0 = x_0)= \frac{\partial}{\partial x_0} C_{Y_{jg},X_0}\bigl( F _{Y_{jg}}(y_{jg}) , x_0\bigr).
$$
For shorthand notation, we let $C_{Y_{jg}|X_0}\bigl( F _{Y_{jg}}(y_{jg})|x_0\bigr) = \frac{\partial}{\partial x_0} C_{Y_{jg},X_0}\bigl( F _{Y_{jg}}(y_{jg}) , x_0\bigr)$. 

The observed variables also load  on the group-specific factors,  hence to account for this dependence, we let $C_{Y_{jg},X_g|X_0 }$ be a bivariate copula that links the observed variable $Y_{jg}$with the group-specific factor $X_g$ given the common factor $X_0$. Hence,
\begin{align*}
& \Pr(Y_{jg} \leq y_{jg}| X_0 =x_0, X_g=x_g)= \frac{\partial}{\partial x_g} \Pr(Y_{jg} \leq y_{jg} , X_g\leq x_g | X_0 =x_0) \\
&= \frac{\partial}{\partial x_g} C_{Y_{jg},X_g|X_0}\bigl(F_{Y_jg|X_0}(y_{jg}|x_0),  x_g\bigr)= C_{Y_{jg}|X_g;X_0}\bigl(F_{Y_jg|X_0}(y_{jg}|x_0)| x_g\bigr).
\end{align*}

 To this end, the pmf of the  bi-factor copula model takes the form
\begin{align}\label{bifactor-pmf}
\pi(\y)&= \int_{[0,1]^{G+1}}  \prod_{g=1}^{G}  \prod_{j=1}^{d_g}  \biggr\{C_{Y_{jg}|X_g;X_0}\bigl(F_{Y_{jg}|X_0}(y_{jg}|x_0)| x_g\bigr) -C_{Y_{jg}|X_g;X_0}\bigr(F_{Y_{jg}|X_0}(y_{jg}-1|x_0)| x_g\bigl) \biggl\} d{x_1}\cdots d{x_G} d{x_0} \nonumber\\
& = \int_0^1  \prod_{g=1}^{G} \Bigg\{ \int_0^1  \prod_{j=1}^{d_g} \biggl[ C_{Y_{jg}|X_g;X_0}\bigl(F_{Y_{jg}|X_0}(y_{jg}|x_0)| x_g\bigr) -C_{Y_{jg}|X_g;X_0}\bigr(F_{Y_{jg}|X_0}(y_{jg}-1|x_0)| x_g\bigl) \biggr] d{x_g} \Bigg\} d{x_0} \nonumber\\
& = \int_0^1  \prod_{g=1}^{G} \Bigg\{ \int_0^1 \prod_{j=1}^{d_g} \biggr[ C_{Y_{jg}|X_g;X_0}\bigl(   C_{Y_{jg}|X_0}(a_{jg,y_{jg}+1}|x_0)   | x_g\bigr)  -C_{Y_{jg}|X_g;X_0}\bigr(    C_{Y_{jg}|X_0}(a_{jg,y_{jg}}|x_0)     | x_g\bigr) \biggl] d{x_g} \Bigg\} d{x_0}\nonumber\\
&= \int_0^1  \prod_{g=1}^{G} \Bigg\{ \int_0^1  \prod_{j=1}^{d_g} 
f_{Y_{jg}|X_{g};X_0}(y_{jg}|x_g,x_0)  d{x_g} \Bigg\} d{x_0}.
\end{align}

\noindent It is shown that the pmf is represented as an one-dimensional integral of a function which is in turn is a product of $G$ one-dimensional integrals. Thus we avoid  $(G+1)$-dimensional numerical integration.  

For the parametric version of the bi-factor copula model, we let $C_{Y_{jg},X_0}$ and  $C_{Y_{jg},X_g|X_0}$ be parametric copulas with dependence parameters $\theta_{jg}$ and  $\delta_{jg}$, respectively.

\subsection{\label{2ndorder-subsec}Second-order copula model}

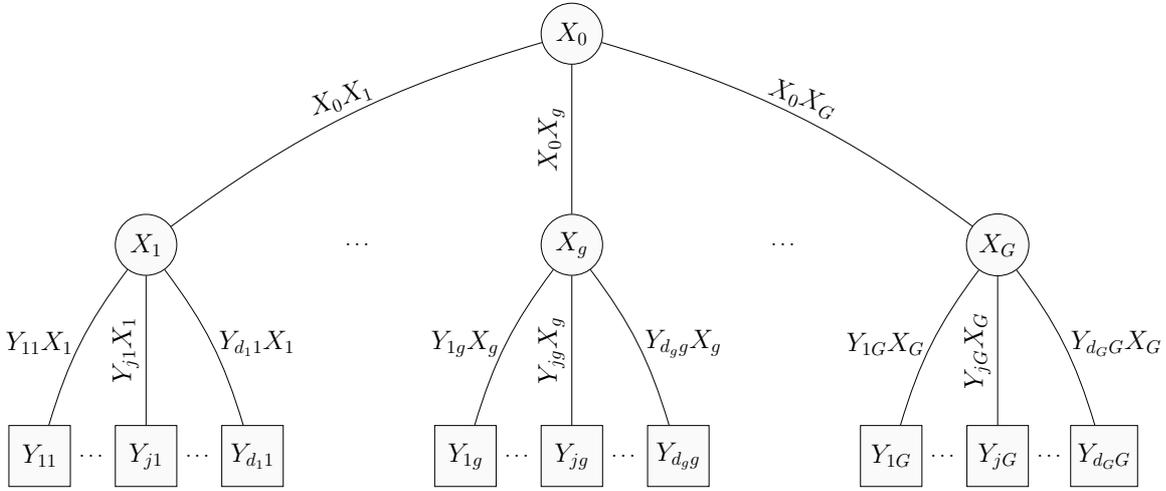
\begin{figure}[h!]
\scalebox{.7}{
    \begin{tikzpicture}
        [square/.style={
            draw,
            fill=gray!20,
            minimum width=3em,
            minimum height=3em,
            node contents={#1}}
            ]

        \node at (0,1) [circle,draw,fill=gray!20,minimum width=3em] (X0) {\Large $X_0$};
       
    		\node at (-8,-3) [circle,draw,fill=gray!20,minimum width=3em] (X1) {\Large $X_1$};
    		        
    		\node at (0,-3) [circle,draw,fill=gray!20,minimum width=3em] (Xg) {\Large $X_g$};

    		\node at (8,-3) [circle,draw,fill=gray!20,minimum width=3em] (XG) {\Large $X_G$};

	    \node at (-10,-7) (Y11) [square={\Large $Y_{11}$}];
         		\node at  (-9,-7)  {$\cdots$};
        \node at (-8,-7) (Yj1) [square={\Large $Y_{j1}$}];
                  \node at  (-7,-7)  {$\cdots$};
        \node at (-6,-7) (Yd11) [square={\Large $Y_{d_11}$}];
\node at  (-4,-3)  {$\cdots$};
	
        \node at (-2,-7) (Y1g) [square={\Large $Y_{1g}$}];
         		\node at  (-1,-7)  {$\cdots$};
        \node at (0,-7) (Yjg) [square={\Large $Y_{jg}$}];
                  \node at  (1,-7)  {$\cdots$};
        \node at (2,-7) (Ydgg) [square={\Large $Y_{d_gg}$}];
\node at  (4,-3)  {$\cdots$};
	
        \node at (6,-7) (Y1G) [square={\Large $Y_{1G}$}];
         		\node at  (7,-7)  {$\cdots$};
        \node at (8,-7) (YjG) [square={\Large $Y_{jG}$}];
                  \node at  (9,-7)  {$\cdots$};
        \node at (10,-7) (YdGG) [square={\Large $Y_{d_GG}$}];
        
        \path(X0) edge[black,bend right=10] node[above,midway,pos=0.5,rotate=20]{\Large $X_0X_1$} (X1);
        \path(X0) edge[black] node[above,midway,pos=0.5,rotate=90]{\Large $X_0X_g$} (Xg);	  
        \path(X0) edge[black,bend left=10] node[above,midway,rotate=-20]{\Large $X_0X_G$}  (XG);	
        
        \path(X1) edge[black,bend right=10] node[left,midway,pos=0.5]{\Large $Y_{11}X_1$} (Y11);	    
        \path(X1) edge[black] node[above,midway,pos=0.5,rotate=90]{\Large $Y_{j1}X_1$} (Yj1);	    
        \path(X1) edge[black,bend left=10] node[right,midway,pos=0.5]{\Large $Y_{d_11}X_1$} (Yd11);	    
        \path(Xg) edge[black,bend right=10] node[left,midway,pos=0.5]{\Large $Y_{1g}X_g$} (Y1g);	    
        \path(Xg) edge[black] node[above,midway,pos=0.5,rotate=90]{\Large $Y_{jg}X_g$} (Yjg);	   
        \path(Xg) edge[black,bend left=10] node[right,midway,pos=0.5]{\Large $Y_{d_gg}X_g$} (Ydgg);	    
        \path(XG) edge[black,bend right=10] node[left,midway,pos=0.5]{\Large $Y_{1G}X_G$} (Y1G);	    
        \path(XG) edge[black] node[above,midway,pos=0.5,rotate=90]{\Large $Y_{jG}X_G$} (YjG);	   
        \path(XG) edge[black,bend left=10] node[right,midway,pos=0.5]{\Large $Y_{d_GG}X_G$} (YdGG); 	  
    \end{tikzpicture}
}

\caption{\label{2ndorder-graph}Graphical representation of the second-order copula model with $G$ first-order 
factors and one second-order  
factor $X_0$.}\label{fig:nested}
\end{figure}

Assume that for a fixed $g=1,\ldots,G$, the items $Y_{1g},\ldots, Y_{d_gg}$ are conditionally independent given the first-order factors $X_g \sim U(0,1),\,g=1,\ldots,G$
and  that $\X=(X_1,\cdots,X_G)$ are conditionally independent given the second-order factor $X_0 \sim U(0,1)$.  That is  the  joint distribution  of $\X$ has an one-factor structure.   We also assume that $Y_{jg}$ in group $g$ does not depend on $X_{g'}$ for $g\neq g'$. 
 Figure \ref{2ndorder-graph} depicts the graphical representation of the model.

The joint pmf takes the form
$$\pi(\y)
 =\int_{[0,1]^G} \Bigg\{    \prod_{g=1}^{G} \prod_{j=1}^{d_g}   \Pr(Y_{jg} = y_{jg} | X_g = x_g) \Bigg\}  c_\X(x_1,\ldots, x_G)  d{x_1}\cdots d{x_G};
$$
$c_\X$ is the one-factor   copula density \citep{Krupskii&Joe-2013-JMVA} of $\X=(X_1,\ldots,X_G)$, viz. 
$$c_\X(x_1,\ldots, x_G) = \int_0^1 \prod_{g=1}^{G} c_{X_g,X_0}(x_g,x_0) dx_0,$$
where $c_{X_g,X_0}$ is the bivariate copula density of the copula $C_{X_g,X_0}$ linking $X_g$ and $X_0$. 

Letting $C_{Y_{jg},X_g}$ be a bivariate copula that joins the observed variable  $Y_{jg}$ and the group-specific factor $X_g$ such that
$$F_{Y_{jg}|X_g}(y_{jg}|x_g):= \Pr(Y_{jg} \leq y_{jg} | X_g = x_g)= \frac{\partial}{\partial x_g} C_{Y_{jg},X_g}\bigl(F _{Y_{jg}}(y_{jg}), x_g\bigr)=C_{Y_{jg}|X_g}\bigl(F _{Y_{jg}}(y_{jg}) | x_g\bigr),$$  
the pmf of the second-order copula model becomes

\begin{small}
\begin{align}\label{2ndorder-pmf}
\pi(\y) & = \int_0^1 \int_{[0,1]^G}  \Bigg\{  \prod_{g=1}^{G} \prod_{j=1}^{d_g}  \Big( C_{Y_{jg}|X_g}\bigl(F _{Y_{jg}}(y_{jg}) | x_g\bigr)  - C_{Y_{jg}|X_g}\bigl(F _{Y_{jg}}(y_{jg}-1) | x_g \bigr) \Big)  \Bigg\} \Bigg\{ \prod_{g=1}^{G}  c_{X_g,X_0}\bigl(x_g,x_0\bigr)  \Bigg\}  d{x_1} \cdots d{x_G} d{x_0}\nonumber \\
& = \int_0^1 \Bigg\{  \prod_{g=1}^{G} \int_0^1 \Bigg[ \prod_{j=1}^{d_g}  \Big( C_{Y_{jg}|X_g}\bigl(F _{Y_{jg}}(y_{jg}) | x_g\bigr) - C_{Y_{jg}|X_g}\bigl(F _{Y_{jg}}(y_{jg}-1) | x_g \bigr) \Big) \Bigg]  c_{X_g,X_0}\bigl(x_g,x_0\bigr)  d{x_g} \Bigg\} d{x_0}\nonumber\\
& = \int_0^1 \Bigg\{  \prod_{g=1}^{G} \int_0^1 \Bigg[ \prod_{j=1}^{d_g}  \Big( C_{Y_{jg}|X_g}\bigl(a_{jg,y_{jg}+1} | x_g\bigr) - C_{Y_{jg}|X_g}\bigl(a_{jg,y_{jg}} | x_g \bigr) \Big) \Bigg]  c_{X_g,X_0}\bigl(x_g,x_0\bigr)  d{x_g} \Bigg\} d{x_0}\nonumber\\
& =\int_0^1  \Bigg\{ \prod_{g=1}^{G}  \int_0^1  \Big[ \prod_{j=1}^{d_g} 
f_{Y_{jg}|X_g}(y_{jg}|x_g) \Big] c_{X_g,X_0}\bigl(x_g,x_0\bigr)  d{x_g} \Bigg\} d{x_0}.
\end{align} 
\end{small} 
Similarly with the bi-factor copula model, the pmf is represented as an one-dimensional integral of a function which is in turn is a product of $G$ one-dimensional integrals. 

For the parametric version of the second-order copula model, we let $C_{Y_{jg},X_g}$ and $C_{X_g,X_0}$ be parametric copulas with dependence parameters $\theta_{jg}$ and $\delta_{g}$, respectively.

\subsection{\label{special-subsec}Special cases}
In this subsection we show what happens when all bivariate copulas are BVN. 
Let $Z_{jg}$ be the underlying continuous variable of the ordinal variable $Y_{jg}$, i.e., 
$Y_{jg} = y_{jg}$ if $\alpha_{jg,y_{jg}} \leq Z_{jg} \leq \alpha_{jg,y_{jg}+1}$
with $\alpha_{jg,K} = \infty$ and $\alpha_{jg,0} = -\infty$.

For the bi-factor model,  if $C_{Y_{jg},X_0}(\cdot;\th_{jg})$ and $C_{Y_{jg},X_g|X_0}(\cdot;\de_{jg})$ are BVN copulas, 
$$ C_{Y_{jg}|X_g; X_0}(C_{Y_{jg} | X_0}(F_{jg}(y_{jg}) | x_0) | x_g) =\Phi  \left( \frac{ \alpha_{jg,y_{jg}+1} - \theta_{jg} \Phi^{-1}(x_0) - \de_{jg} \sqrt{1-\theta_{jg}^2} \Phi^{-1}(x_g)}{\sqrt{(1 - \theta_{jg}^2) (1-\de_{jg}^2) }}   \right).$$
Hence, the pmf for the bi-factor copula model in (\ref{bifactor-pmf}) becomes

\begin{eqnarray*}
\pi(\y)&=& \int_0^1 \prod_{g=1}^G \Biggl\{ \int_0^1  \prod_{j=1}^{d_g}  \Biggl[ \Phi  \left( \frac{ \alpha_{jg,y_{jg}+1} - \theta_{jg} \Phi^{-1}(x_0) - \de_{jg} \sqrt{1-\theta_{jg}^2} \Phi^{-1}(x_g)}{\sqrt{(1 - \theta_{jg}^2) (1-\de_{jg}^2) }}   \right)    -   \\&&\Phi  \left( \frac{ \alpha_{jg,y_{jg}} - \theta_{jg} \Phi^{-1}(x_0) - \de_{jg} \sqrt{1-\theta_{jg}^2} \Phi^{-1}(x_g)}{\sqrt{(1 - \theta_{jg}^2) (1-\de_{jg}^2) }}   \right)  \Biggr]   dx_g \Biggr\} dx_0  \\
 & =& \int_{-\infty}^{\infty}  \prod_{g=1}^G \Biggl\{  \int_{-\infty}^{\infty} \prod_{j=1}^{d_g} \Biggl[ \Phi  \left( \frac{ \alpha_{jg,y_{jg}+1} - \theta_{jg} z_{0} - \de_{jg} \sqrt{1-\theta_{jg}^2} z_{g}}{\sqrt{(1 - \theta_{jg}^2) (1-\de_{jg}^2) }}      \right) 
\\&&- \Phi  \left( \frac{ \alpha_{jg,y_{jg}} - \theta_{jg} z_{0} - \de_{jg} \sqrt{1-\theta_{jg}^2} z_{g}}{\sqrt{(1 - \theta_{jg}^2) (1-\de_{jg}^2) }}      \right)\Biggr]    \phi(z_{g}) dz_{g}   \Biggr\} \phi(z_{0})  dz_{0}.
\end{eqnarray*}
This model is the same as the bi-factor Gaussian model \citep{Gibbons1992-PKA,Gibbons-etal2007} with stochastic representation 
\begin{equation} \label{eq:bifactorBVN}
Z_{jg} = \theta_{jg} Z_{0} + \gamma_{jg} Z_{g} +\sqrt{1 - \theta_{jg}^2 - \gamma_{jg}^2} \epsilon_{jg},  \qquad g=1,\ldots,G,\quad   j=1,\cdots,d_g, 
\end{equation}
where $\gamma_{jg}=\de_{jg}\sqrt{1 - \theta_{jg}^2}$ and $Z_{0},Z_{g},\epsilon_{jg}$ are iid $N(0,1)$ random variables. The parameter $\theta_{jg}$ of $C_{Y_{jg},X_0}$ is the correlation of $Z_{jg}$ and $Z_0$, and the parameter $\delta_{jg}$ of $C_{Y_{jg},X_g|X_0}$ is the partial correlation between $Z_{jg}$ and $Z_{g}=\Phi^{-1}(X_g)$ given $Z_0=\Phi^{-1}(X_0)$.

It implies that the underlying random variables $Z_{jg}$'s 
have a  multivariate Gaussian distribution where the off-diagonal entries of the correlation matrix have the form  $\theta_{j_1g}\theta_{j_2g} + \gamma_{j_1g}\gamma_{j_2g}$   and $\theta_{j_1g_1}\theta_{j_2g_2}$ for $ j_1 \neq j_2$ and   $g_1 \neq g_2$, respectively. 
For the Gaussian bi-factor model to be identifiable,  the number of dependence parameters has to be $2d - N_{1}-N_{2}$, where $N_{1}$ and $N_{2}$ is the number of groups that consist of 1 and 2 items, respectively. For a group g of size 1 with variable $j$, $Z_g$ is absorbed with $\epsilon_{jg}$ because $\gamma_{jg}$ would not be identifiable. For a group $g$ of size 2 with variable indices $j_1,j_2$, the parameters $\gamma_{j_1g}$ and $\gamma_{j_2g}$ appear only in one correlation, hence 
one of $\gamma_{j_1g},\gamma_{j_2g}$ can be taken as 1 without loss of generality. For the bi-factor copula with non-Gaussian linking copulas, near non-identifiability can occur when there are groups of size 2; in this case, one of the linking copulas to the group latent variable can be fixed  at comonotonicity.

For the Gaussian second-order model let 
$Z_0,Z_1',\ldots,Z_G'$ be  the dependent latent $N(0,1)$ variables, where $Z_0$ is the second-order factor and $Z_g'=\beta_{g}Z_0+(1-\beta_{g}^2)Z_g$ is the first-order factor for group $g$. That is, there is an one second-order factor $Z_0$, and the first-order factors $Z_1',\ldots,Z_G'$ are linear combinations of the second-order factor, plus a unique variable $Z_g$ for each first-order factor.  
The stochastic representation is  \citep{Krupskii&Joe-2015-JMVA}:
\begin{eqnarray*}
Z_{jg} &=& \beta_{jg} Z _g' +\sqrt{1-\beta_{jg}^2}\epsilon_{jg}\\
Z_g'&=&\beta_{g}Z_0+\sqrt{1-\beta_{g}^2}Z_g, \qquad g=1,\ldots,G,  \quad j=1,\cdots,d_g, 
\end{eqnarray*}
or 
\begin{equation}\label{2nd-order-gaussian}
Z_{jg}=\beta_{jg} \beta_{g}Z_0+\beta_{jg}\sqrt{1-\beta_{g}^2}Z_g +\sqrt{1-\beta_{jg}^2}\epsilon_{jg},  \quad j=1,\cdots,d_g.
\end{equation}
Hence, this is a special case of  (\ref{eq:bifactorBVN})
where $\theta_{jg} = \beta_{jg}  \beta_{g}$ and $\gamma_{jg} = \beta_{jg} \sqrt{1-\beta_{g}^2}$. 

\baselineskip=24pt

\section{\label{sec-computing}Estimation and computational details}

For the set of all  parameters, let   $\thbf=(\bfa,\thbf_g,\debf_{g})$ for the bi-factor copula model and   $\thbf=(\bfa,\thbf_g,\debf)$  for the second-order copula model, where $\bfa=(a_{jg,k}: j=1,\ldots,d_g, g=1,\ldots,G,  k=1,\ldots,K-1)$, 
$\thbf_g=(\theta_{1g}, \ldots, \theta_{jg}, \ldots,\theta_{d_gg}: g=1,\ldots,G)$, $\debf_{g}=(\de_{1g}, \ldots, \de_{jg}, \ldots,\de_{d_gg}: g=1,\ldots,G)$  and $\debf=(\de_{1}, \ldots,\de_{G})$.

With sample size $n$ and data $\y_1,\ldots,\y_n$, the  joint log-likelihood of the bi-factor and second-order copula is 
\begin{equation}\label{joint-loglik}
\ell(\thbf;\y_1,\ldots,\y_n)=\sum_{i=1}^n\log \pi (\y_i;\thbf).
\end{equation}
with $\pi (\y_i;\thbf)$ as in (\ref{bifactor-pmf}) and (\ref{2ndorder-pmf}), respectively. Maximization of (\ref{joint-loglik})
is numerically possible
but is time-consuming  for large $d$ because of many univariate cutpoints and
dependence parameters.
Hence, we approach estimation using the  two-step IFM method proposed  by \cite{Joe2005-JMVA} that 
can efficiently, in the sense of computing time and asymptotic variance,
estimate the model parameters.

 In the first step, the cutpoints are estimated using the univariate sample proportions. The univariate cutpoints for the $j$th item in group $g$ are estimated as $\hat{a}_{jg,k} = \sum_{y=0}^{k} p_{jg,y}$, where  $p_{jg,y}\,,y=0,\ldots,K-1$ for $g=1,\ldots,G$ and  $j=1,\ldots,d_g$ are the univariate sample proportions. 
In the second step of the IFM method, the joint log-likelihood in (\ref{joint-loglik}) is maximized over the copula parameters with the cutpoints fixed as estimated at the first step. The estimated copula parameters can be obtained by using a quasi-Newton \citep{Nash1990} method applied to the logarithm of the joint likelihood.

For the bi-factor  copula model numerical evaluation of the joint pmf can be achieved with the following steps:
\begin{enumerate}
\itemsep=10pt
\item Calculate Gauss-Legendre quadrature \citep{Stroud&Secrest1966} points  $\{x_q: q=1,\ldots,n_q\}$  and weights $\{w_q: q=1,\ldots,n_q\}$  in terms of standard uniform.  

\item Numerically evaluate the joint pmf 
$$\int_0^1  \prod_{g=1}^{G} \Bigg\{ \int_0^1  \prod_{j=1}^{d_g} 
f_{Y_{jg}|X_{jg};X_0}(y_{jg}|x_g,x_0)  d{x_g} \Bigg\} d{x_0}$$

in a double sum
$$\sum_{q_1=1}^{n_q}  w_{q_1} \prod_{g=1}^{G} \Bigg\{ \sum_{q_2=1}^{n_q} w_{q_2} \prod_{j=1}^{d_g} 
f_{Y_{jg}|X_{jg};X_0}(y_{jg}|x_{q_2},x_{q_1})  \Bigg\} $$

\end{enumerate}

For the second-order  copula model numerical evaluation of the joint pmf can be achieved with the following steps:

\begin{enumerate}
\itemsep=10pt
\item Calculate Gauss-Legendre quadrature points $\{x_q : q=1,\ldots, n_q \}$ and weights $\{w_q : q=1,\ldots, n_q \}$ in terms of stand uniform.
\item Numerically evaluate the joint pmf

$$\int_0^1  \Bigg\{ \prod_{g=1}^{G}  \int_0^1  \Big[ \prod_{j=1}^{d_g} 
f_{Y_{jg}|X_g}(y_{jg}|x_g;\theta_{jg}) \Big] c_{X_g,X_0}\bigl(x_g,x_0;\de_{g}\bigr)  d{x_g} \Bigg\} d{x_0}$$

in a double sum 
$$\sum_{q_1=1}^{n_q}   w_{q_1}  \Bigg\{ \prod_{g=1}^{G}  \sum_{q_2=1}^{n_q}  w_{q_2} \Big[ \prod_{j=1}^{d_g} 
f_{Y_{jg}|X_g}(y_{jg}|x_{q_2|q_1};\theta_{jg}) \Big]  \Bigg\} $$
where $x_{q_2|q_1} = C^{-1}_{Y_{jg}|X_g;X_0}( x_{q_2} | x_{q_1};\de_{g})$. 
Note that the independent quadrature
points $\{x_{q_1}: q_1 = 1,\ldots,n_q\}$ and  $\{x_{q_2}: q_2 = 1,\ldots,n_q\}$ have converted to dependent quadrature points that have an one-factor copula distribution $C_{X}(\cdot;\debf)$. 
\end{enumerate}
With Gauss-Legendre quadrature, the same nodes and weights are used for different functions; this helps in
yielding smooth numerical derivatives for numerical optimization via quasi-Newton. Our comparisons show that $n_q=25$ quadrature points are adequate with good precision.

\section{\label{sec-selection}Bivariate copula selection}
In line with \cite{Nikoloulopoulos2015-PKA}, we use bivariate parametric copulas  that can be used when considering latent maxima, minima or mixtures of means, namely the Gumbel, survival Gumbel (s.Gumbel) and Student $t_\nu$ copulas, respectively. A model with bivariate  Gumbel copulas that possess upper tail dependence has latent (ordinal) variables that can be considered as
(discretized) maxima, and there is more probability in the joint upper tail.  A model with bivariate s.Gumbel copulas that possess  lower tail dependence has latent (ordinal) variables that can be considered as
(discretized) minima, and there is more probability in the joint lower tail. A model with bivariate $t_\nu$ copulas  that possess the same lower and upper tail dependence has latent (ordinal) variables that can be considered as mixtures of (discretized) means, since the bivariate Student $t_\nu$ distribution arises as a scale mixture of bivariate normals. A small value of $\nu$, such as $1 \leq \nu\leq 5$, leads to a model with more probabilities in the joint upper and joint lower tails compared with the BVN copula.

In the following subsections we describe  simple diagnostics based on semi-correlations and an heuristic method that automatically selects the bivariate parametric copula families  that build either the bi-factor or the second-order copula model.  
In the context of items  that can be split into $G$ non-overlapping  groups, such that there is homogeneous dependence within each group, it is sufficient  to (a)  summarize the average of the polychoric semi-correlations   for all pairs within each of the $G$ groups and for all pairs of items, and (b)   not mix bivariate copulas for a single factor; hence, for both the bi-factor  and second-order copula models we allow $G+1$ different copula families, one for each group specific factor $X_g$ and one for $X_0$.

\subsection{\label{sec:semicor}Simple diagnostics based on  semi-correlations}

Choices of copulas with upper or lower tail dependence are better if the items have more probability in joint lower or upper tail than would be expected with the BVN copula.
This can be shown with summaries of correlations  in the upper joint tail and lower joint tail.

Consider again the underlying $N(0,1)$ latent variables $Z_{jg}$'s of the ordinal variables $Y_{jg}$'s. The correlations of  $Z_{jg}$'s in the upper and lower tail, hereafter semi-correlations, are defined as \citep[page 71]{Joe2014-CH}:
\begin{eqnarray}\label{semicor}
\rho_N^+&=&\mbox{Cor}\Bigl(Z_{j_1g},Z_{j_2g}|Z_{j_1g}>0,Z_{j_2g}>0\Bigr)\\
&=&\frac{\int_{0}^{\infty}\int_{0}^{\infty} z_1z_2\phi(z_1)\phi(z_2)c\bigl(\Phi(z_1),\Phi(z_2)\bigr)dz_1dz_2-{\biggr(\int_0^\infty z\phi(z)\Bigr(1- C_{2|1}\bigr(0.5|\Phi(z)\bigl)\Bigr)dz\biggr)^2}/{C(0.5,0.5)}}
{\int_0^\infty z^2\phi(z)\Bigr(1- C_{2|1}\bigr(0.5|\Phi(z)\bigl)\Bigr)dz-{\biggr(\int_0^\infty z\phi(z)\Bigr(1- C_{2|1}\bigr(0.5|\Phi(z)\bigl)\Bigr)dz\biggr)^2}/{C(0.5,0.5)}};\nonumber \\
\rho_N^-&=&\mbox{Cor}\Bigl(Z_{j_1g},Z_{j_2g}|Z_{j_1g}<0,Z_{j_2g}<0\Bigr)\nonumber\\
&=&\frac{\int_{-\infty}^{0}\int_{-\infty}^{0} z_1z_2\phi(z_1)\phi(z_2)c\bigl(\Phi(z_1),\Phi(z_2)\bigr)dz_1dz_2-{\biggr(\int_{-\infty}^0 z\phi(z)C_{2|1}\bigr(0.5|\Phi(z)\bigl)dz\biggr)^2}/{C(0.5,0.5)}}
{\int_{-\infty}^0 z^2\phi(z) C_{2|1}\bigr(0.5|\Phi(z)\bigl)dz-{\biggr(\int_{-\infty}^0 z\phi(z)C_{2|1}\bigr(0.5|\Phi(z)\bigl)dz\biggr)^2}/{C(0.5,0.5)}}.\nonumber
\end{eqnarray}

From the above expressions, it is clear that the  semi-correlations depend only on the copula C of $\Bigl(\Phi(Z_{j_1g}),$ $\Phi(Z_{j_2g})\Bigr)$; $C_{2|1}$ is the conditional copula cdf. 
For the BVN and $t_\nu$ copulas  $\rho_N^{-}=\rho_N^{+}$, while for the Gumbel and s.Gumbel copulas $\rho_N^{-}<\rho_N^{+}$ and   $\rho_N^{-}>\rho_N^{+}$, respectively. 
The sample versions of $\rho_N^{+},\rho_N^{-}$  for item response data are the polychoric correlations  in the joint lower and upper  quadrants of $Y_{j_1g}$ and $Y_{j_2g}$ \citep{Kadhem&Nikoloulopoulos2019}.

\subsection{Selection algorithm} \label{sec:selecalg}

We propose an heuristic method that  selects appropriate bivariate copulas for each factor of the bi-factor and second-order copula  models. It starts with an initial assumption, that all bivariate linking copulas are BVN copulas, i.e. the starting model is either the Gaussian bi-factor or second-order model, and then sequentially other copulas with lower or upper tail dependence are assigned to the factors where necessary to account for more probability in one or both joint tails.  
The selection algorithm involves the following steps:

\begin{enumerate}
\itemsep=10pt
\item Fit the bi-factor or second-order copula model  with BVN copulas.

\item Fit all the possible bi-factor or second-order copula models, iterating over all the copula candidates that link all items $Y_{jg}$'s in group $g$ or  each group-specific factor $X_g$, respectively, to  $X_0$.  

\item Select the copula family that corresponds to the lowest Akaike information criterion (AIC), that is, $\text{AIC}= -2 \times \ell +2 \times \#\text{copula parameters}$.

\item Fix the selected copula family  that links the observed (bi-factor model) or  latent (second-order model) variables  to $X_0$.

\item For $g=1,\dots,G$: 
\begin{enumerate}
\itemsep=10pt
\item Fit all the possible models, iterating over all the copula candidates that link all the items in group $g$  to the group-specific factor $X_g$.
\item Select the copula family that corresponds to the lowest AIC.
\item Fix the selected linking copula family for all the items in group $g$ with $X_g$.

\end{enumerate}
\end{enumerate}

\section{\label{sec-gof}Goodness-of-fit}
We will use the limited information $M_2$ statistic proposed by \cite{Maydeu-Olivares&Joe2006} to evaluate the overall fit of the proposed bi-factor and second-order  copula models. 
The $M_2$ statistic is based on a quadratic form of the deviations of sample and model-based proportions over all bivariate margins. 
For our parametric  models with parameter vector $\thbf$ of dimension $q$, let $\pibf_2(\thbf)=\bigl(\dot{\pibf}_1(\thbf)^\top,\dot{\pibf}_2(\thbf)^\top\bigr)^\top$ be the column vector of the
 univariate  and bivariate  model-based  marginal probabilities that do not include category 0  with sample counterpart 
$\pp_2=(\dot{\pp}_1^\top,\dot{\pp}_2^\top)^\top$. 
The total number of the univariate and bivariate residuals $\bigl(\pp_2-\pibf_2(\hat\thbf)\bigr)^\top$ is
\begin{align*}
s = d (K-1)  +  \binom{d}{2} (K - 1)^2,
\end{align*}
where $d (K-1)$ is the dimension of the univariate residuals and $\binom{d}{2} (K - 1)^2$ is the dimension of the bivariate residuals excluding category 0.

With a sample size $n$, 
the limited-information $M_2$  statistic 
is given by
\begin{equation}\label{M_2}
M_2=M_2(\hat\thbf)=n\bigl(\pp_2-\pibf_2(\hat\thbf)\bigr)^\top \C_2(\hat\thbf)\bigl(\pp_2-\pibf_2\bigl(\hat\thbf)\bigr),
\end{equation}
with
\begin{equation}\label{C_2}
\C_2(\thbf)=\Xibf_2^{-1}-\Xibf_2^{-1}\Debf_2(\Debf_2^\top\Xibf_2^{-1}\Debf_2)^{-1}\Debf_2^\top\Xibf_2^{-1}
=\Debf_2^{(c)}\bigl([\Debf_2^{(c)}]^\top\Xibf_2\Debf_2^{(c)}\bigr)^{-1}[\Debf_2^{(c)}]^\top,
\end{equation}
where $\Debf_2=\partial\pibf_2(\thbf)/\partial\thbf^\top$ is an $s\times q$ matrix   with   the first order derivatives of the univariate and bivariate marginal probabilities with respect to the estimated model parameters (in the Appendix, we provide details on the calculation of these derivatives), $\Debf_2^{(c)}$ is an $s\times(s-q)$ orthogonal complement to $\Debf_2$,
such that $[\Debf_2^{(c)}]^\top\Debf_2=\mathbf{0}$, and  $\Xibf_2$ is the  asymptotic $s \times s$ covariance matrix of $\sqrt{n}\bigl(\pp_2-\pibf_2(\hat\thbf)\bigr)^\top$. The limited information statistic $M_2$ under the null hypothesis has an asymptotic
distribution that is $\chi^2$ with $s-q$ degrees of freedom
when the estimate $\hat \thbf$ is $\sqrt{n}$-consistent. 

The asymptotic covariance matrix $\Xibf_2$ can 
be partitioned according to the portioning of $\pp_2$ into $\Xibf_{11}=\sqrt{n}\mbox{Acov}(\dot{\pp}_1)$, $\Xibf_{21}=\sqrt{n}\mbox{Acov}(\dot{\pp}_2,\dot{\pp}_1)$ and $\Xibf_{22}=\sqrt{n}\mbox{Acov}(\dot{\pp}_2)$, where $\mbox{Acov}(\cdot)$ denotes asymptotic covariance matrix. The elements of $\Xibf_{11}$, $\Xibf_{21}$ and $\Xibf_{22}$ involve up to the 4-dimensional probabilities as shown below: 
\begin{eqnarray*}
\sqrt{n}\mbox{Acov}(p_{j_1,y_1},p_{j_2,y_2})&=&\pi_{j_1j_2,y_1y_2}-\pi_{j_1,y_1}\pi_{j_2,y_2}\\
\sqrt{n}\mbox{Acov}(p_{j_1j_2,y_1y_2},p_{j_3,y_3})&=&\pi_{j_1j_2j_3,y_1y_2y_3}-\pi_{j_1j_2,y_1y_2}\pi_{j_3,y_3}\\
\sqrt{n}\mbox{Acov}(p_{j_1j_2,y_1y_2},p_{j_3j_4,y_3y_4})&=&\pi_{j_1j_2j_3j_4,y_1y_2y_3y_4}-\pi_{j_1j_2,y_1y_2}\pi_{j_3j_4,y_3y_4},
\end{eqnarray*}
where
$
\pi_{j,y} = \Pr(Y_j = y)$, 
$\pi_{j_1j_2,y_1y_2}= \Pr(Y_{j_1} = y_1,Y_{j_2} = y_2)$, 
$\pi_{j_1j_2j_3,y_1y_2y_3}=\Pr(Y_{j_1} = y_1,Y_{j_2} = y_2,Y_{j_3} = y_3)$, and 
$\pi_{j_1j_2j_3j_4,y_1y_2y_3y_4}=\Pr(Y_{j_1} = y_1,Y_{j_2} = y_2,Y_{j_3} = y_3,Y_{j_4} = y_4)$.

\section{\label{sec-sim}Simulations}
An extensive simulation study is conducted to 
(a) gauge the small-sample efficiency of the IFM estimation  method and investigate the misspecification of the  bivariate pair-copulas, (b) examine the reliability of using the heuristic  algorithm to select the true (simulated)   bivariate linking copulas, and (c)  study the small-sample performance of the  $M_2$ statistic.

\begin{sidewaystable}[htbp]
  \centering
  \small
       \setlength{\tabcolsep}{6pt}  
  \caption{\label{tab:misspecified} Small sample of size $n = 500$ simulations (10$^3$ replications) from  the bi-factor and second-order factor models  with  Gumbel copulas and group estimated average
 biases, root mean square errors (RMSE), and standard deviations (SD), scaled by $n$, for the IFM estimates  under different pair-copulas from  the bi-factor and second-order  copula  models.}
    \begin{tabular}{llcccccccccccrccccccccc}
    \toprule
      &   &   &   & \multicolumn{9}{c}{Bi-factor copula model} &   & \multicolumn{9}{c}{Second-order copula model} \\
\cmidrule{5-13}\cmidrule{15-23}      &   &   &   & \multicolumn{4}{c}{$\tau(\thbf_g),\,g=1,\dots,4$} &   & \multicolumn{4}{c}{$\tau(\debf_g),\,g=1,\dots,4$}&   & \multicolumn{4}{c}{$\tau(\debf)$} &   & \multicolumn{4}{c}{$\tau(\thbf_g),\,g=1,\dots,4$} \\
\cmidrule{5-8}\cmidrule{10-13}\cmidrule{15-18}\cmidrule{20-23}      & \multicolumn{1}{c}{Fitted model} & K &   & 0.45 & 0.55 & 0.65 & 0.75 &   & 0.30 & 0.35 & 0.40 & 0.50 &   & 0.30 & 0.35 & 0.40 & 0.45 &   & 0.40 & 0.50 & 0.60 & 0.70 \\
\cmidrule{1-13}\cmidrule{15-23}    $n$bias & BVN & 3 &   & 2.65 & 2.54 & 2.66 & 2.16 &   & 6.60 & 7.81 & 6.99 & 6.39 &   & 5.58 & 5.34 & 5.33 & 5.60 &   & 0.41 & 0.86 & 0.62 & 0.27 \\
      &   & 5 &   & 1.98 & 2.27 & 2.54 & 2.53 &   & 5.99 & 6.27 & 5.42 & 2.31 &   & 8.71 & 8.36 & 7.94 & 8.52 &   & 0.93 & 0.51 & 0.58 & 2.52 \\
      & Gumbel & 3 &   & 0.39 & 0.35 & 0.28 & 0.34 &   & 0.89 & 1.02 & 1.62 & 3.40 &   & -0.18 & 0.18 & 0.18 & 1.88 &   & 0.22 & 0.67 & 1.14 & 2.37 \\
      &   & 5 &   & 0.23 & 0.23 & 0.07 & 0.20 &   & 0.84 & 0.85 & 1.02 & 1.98 &   & 0.22 & 0.13 & -0.25 & 1.15 &   & 0.23 & 0.43 & 0.63 & 0.60 \\
      & s.Gumbel & 3 &   & 3.59 & 3.03 & 1.51 & 0.31 &   & 4.86 & 4.52 & 4.21 & 1.19 &   & 18.43 & 18.29 & 18.54 & 18.68 &   & 6.32 & 6.18 & 5.47 & 3.67 \\
      &   & 5 &   & 0.79 & 2.25 & 3.80 & 5.30 &   & 15.89 & 15.82 & 13.89 & 14.52 &   & 25.65 & 24.80 & 23.58 & 22.59 &   & 3.77 & 2.54 & 1.24 & 2.74 \\
      & $t_5$ & 3 &   & 1.65 & 2.81 & 3.28 & 3.48 &   & 6.99 & 8.20 & 7.07 & 4.89 &   & 7.98 & 8.55 & 9.18 & 9.55 &   & 3.36 & 3.56 & 4.71 & 3.81 \\
      &   & 5 &   & 0.49 & 0.49 & 0.84 & 0.92 &   & 5.81 & 6.09 & 5.58 & 1.69 &   & 9.71 & 10.05 & 9.82 & 9.87 &   & 2.24 & 2.29 & 2.64 & 0.36 \\
    \midrule
    $n$SE & BVN & 3 &   & 15.03 & 13.42 & 12.37 & 11.06 &   & 30.77 & 31.20 & 33.07 & 39.93 &   & 22.80 & 24.94 & 24.97 & 27.03 &   & 16.82 & 16.41 & 17.06 & 21.32 \\
      &   & 5 &   & 13.68 & 11.89 & 10.63 & 8.95 &   & 24.58 & 25.33 & 25.70 & 29.86 &   & 21.28 & 23.04 & 22.45 & 24.72 &   & 15.09 & 14.27 & 14.01 & 15.41 \\
      & Gumbel & 3 &   & 15.10 & 13.81 & 12.33 & 10.97 &   & 29.61 & 31.34 & 32.82 & 42.17 &   & 22.58 & 24.73 & 25.35 & 27.87 &   & 16.99 & 16.73 & 17.66 & 22.02 \\
      &   & 5 &   & 13.67 & 12.29 & 10.55 & 8.76 &   & 23.60 & 24.72 & 25.39 & 31.13 &   & 20.75 & 22.75 & 22.69 & 24.86 &   & 15.31 & 14.62 & 14.33 & 15.72 \\
      & s.Gumbel & 3 &   & 15.58 & 13.76 & 12.60 & 11.27 &   & 33.77 & 34.80 & 38.18 & 51.31 &   & 25.34 & 26.80 & 27.19 & 29.36 &   & 17.40 & 16.49 & 16.59 & 18.46 \\
      &   & 5 &   & 14.11 & 12.30 & 11.16 & 9.66 &   & 27.08 & 28.44 & 30.18 & 40.10 &   & 22.61 & 24.13 & 23.36 & 25.46 &   & 15.90 & 14.57 & 14.38 & 16.89 \\
      & $t_5$ & 3 &   & 15.29 & 13.54 & 12.27 & 10.79 &   & 31.43 & 31.74 & 33.02 & 39.02 &   & 23.59 & 25.57 & 25.65 & 27.61 &   & 17.48 & 16.69 & 17.64 & 22.03 \\
      &   & 5 &   & 13.84 & 11.99 & 10.55 & 8.80 &   & 24.79 & 25.35 & 25.66 & 29.10 &   & 21.67 & 23.52 & 22.67 & 24.52 &   & 15.40 & 14.52 & 14.03 & 14.88 \\
    \midrule
    $n$RMSE & BVN & 3 &   & 15.28 & 13.66 & 12.66 & 11.27 &   & 31.48 & 32.19 & 33.81 & 40.45 &   & 23.47 & 25.50 & 25.53 & 27.60 &   & 16.83 & 16.44 & 17.08 & 21.33 \\
      &   & 5 &   & 13.83 & 12.11 & 10.93 & 9.30 &   & 25.31 & 26.10 & 26.27 & 29.96 &   & 22.99 & 24.51 & 23.81 & 26.14 &   & 15.12 & 14.28 & 14.03 & 15.62 \\
      & Gumbel & 3 &   & 15.10 & 13.81 & 12.34 & 10.98 &   & 29.63 & 31.37 & 32.87 & 42.31 &   & 22.58 & 24.73 & 25.35 & 27.94 &   & 16.99 & 16.75 & 17.71 & 22.15 \\
      &   & 5 &   & 13.67 & 12.30 & 10.55 & 8.77 &   & 23.62 & 24.74 & 25.42 & 31.20 &   & 20.75 & 22.75 & 22.69 & 24.88 &   & 15.31 & 14.63 & 14.35 & 15.73 \\
      & s.Gumbel & 3 &   & 16.00 & 14.09 & 12.69 & 11.27 &   & 34.13 & 35.13 & 38.42 & 51.33 &   & 31.33 & 32.45 & 32.91 & 34.80 &   & 18.52 & 17.65 & 17.49 & 18.82 \\
      &   & 5 &   & 14.14 & 12.51 & 11.79 & 11.02 &   & 31.41 & 32.55 & 33.22 & 42.67 &   & 34.19 & 34.60 & 33.19 & 34.04 &   & 16.35 & 14.82 & 14.44 & 17.13 \\
      & $t_5$ & 3 &   & 15.40 & 13.83 & 12.71 & 11.34 &   & 32.21 & 32.80 & 33.77 & 39.32 &   & 24.91 & 26.97 & 27.24 & 29.21 &   & 17.80 & 17.08 & 18.27 & 22.36 \\
      &   & 5 &   & 13.85 & 12.01 & 10.59 & 8.86 &   & 25.47 & 26.08 & 26.26 & 29.16 &   & 23.75 & 25.58 & 24.71 & 26.43 &   & 15.56 & 14.71 & 14.29 & 14.89 \\
    \bottomrule
    \end{tabular}%
\end{sidewaystable}%

We randomly generate 1,000 datasets with samples of size $n=500$ or 1000 and $d=16$ items, with $K=3$ or $K=5$ equally weighted categories, that are equally separated into  $G=4$ non-overlapping groups from the bi-factor and second-order copula model. In each simulated model, we use  different linking copulas to cover different types of dependence. 
To make 
the models comparable, we convert the BVN/$t_\nu$  and Gumbel/s.Gumbel copula  parameters 
to Kendall's $\tau$'s via  
\begin{equation}\label{tauBVN}
\tau(\th)=\frac{2}{\pi}\arcsin(\th)
\end{equation}
and
\begin{equation}\label{tauGumbel}
\quad \tau(\th)=1-\th^{-1},
\end{equation}
respectively. 
For the bi-factor copula models we set  $\tau(\thbf_g)=(0.45,0.55,0.65,0.75)$ and  $\tau(\debf_g)=(0.30,0.35,$ $0.40,0.50)$ for $g=1,\ldots,4$. For the second-order copula models we set   $\tau(\thbf_g)=(0.4,0.5,0.6,0.7)$ for $g=1,\ldots,4$ and $\tau(\debf)=(0.30,0.35,0.40,0.45)$.

\baselineskip=26pt

 The Kendall's tau parameters $\tau(\thbf_g)$ and $\tau(\debf_g)$ as described above are common for each  group, hence Table \ref{tab:misspecified}  contains the group estimated average
 biases, root mean square errors (RMSE), and standard deviations (SD), scaled by $n$, for the IFM estimates  under different pair-copulas from  the bi-factor and second-order  copula  models. In the true (simulated) models the linking copulas  are Gumbel copulas.

Conclusions from the values in the table are the following:

\begin{itemize}
\item IFM   with  the true bi-factor or  second-order  model is highly efficient according to the simulated biases, SDs and RMSEs.
\item The IFM estimates of $\tau$'s  are not  robust under copula misspecification and  their biases  increase when the assumed bivariate copula  has tail dependence of opposite direction from the true bivariate copula. For example, in Table \ref{tab:misspecified}  the  scaled biases for the IFM estimates  increase substantially  when the linking copulas are the s.Gumbel   copulas.  
\end{itemize}

\begin{table}[!h]
  \centering
  \small
      \setlength{\tabcolsep}{2.7pt}  
  \caption{\label{tab:model-selection} Small sample of size  $n = 500$ simulations ($10^3$ replications) from the bi-factor and second-order factor models with various linking copulas and frequencies of the true bivariate copula identified using the model selection algorithm.}
    \begin{tabular}{lcccccccccccccccc}
    \toprule
    Bi-factor &   & \multicolumn{3}{c}{Model 1 } &   & \multicolumn{3}{c}{Model 2} &   & \multicolumn{3}{c}{Model 3} &   & \multicolumn{3}{c}{Model 4} \\
\cmidrule{3-5}\cmidrule{7-9}\cmidrule{11-13}\cmidrule{15-17}      &   & Copula & $K=3$ & $K=5$ &   & Copula & $K=3$ & $K=5$ &   & Copula & $K=3$ & $K=5$ &   & Copula & $K=3$ & $K=5$ \\
    \midrule
    $X_0$ &   & Gumbel & 992 & 1000 &   & $t_5$ & 984 & 1000 &   & Gumbel & 996 & 1000 &   & $t_5$ & 975 & 1000 \\
    $ X_1$ &   & Gumbel & 858 & 956 &   & $t_5$ & 597 & 806 &   & $t_5$ & 585 & 789 &   & Gumbel & 888 & 958 \\
    $ X_2$ &   & Gumbel & 870 & 951 &   & $t_5$ & 588 & 799 &   & $t_5$ & 569 & 775 &   & Gumbel & 894 & 969 \\
    $ X_3$ &   & Gumbel & 846 & 950 &   & $t_5$ & 546 & 777 &   & s.Gumbel & 844 & 945 &   & s.Gumbel & 865 & 947 \\
    $ X_4$ &   & Gumbel & 844 & 942 &   & $t_5$ & 589 & 805 &   & s.Gumbel & 878 & 949 &   & s.Gumbel & 900 & 956 \\
   
    \end{tabular}%

      \setlength{\tabcolsep}{2.4pt}  
    \begin{tabular}{lcccccccccccccccc}
    \toprule
    \multicolumn{2}{l}{Second-order} & \multicolumn{3}{c}{Model 1 } &   & \multicolumn{3}{c}{Model 2} &   & \multicolumn{3}{c}{Model 3} &   & \multicolumn{3}{c}{Model 4} \\
\cmidrule{3-5}\cmidrule{7-9}\cmidrule{11-13}\cmidrule{15-17}      &   & Copula & $K=3$ & $K=5$ &   & Copula & $K=3$ & $K=5$ &   & Copula & $K=3$ & $K=5$ &   & Copula & $K=3$ & $K=5$ \\
    \midrule
    $X_0$ &   & Gumbel & 901 & 848 &   & $t_5$ & 664 & 819 &   & Gumbel & 892 & 987 &   & $t_5$ & 648 & 765 \\
    $X_1$ &   & Gumbel & 895 & 975 &   & $t_5$ & 735 & 939 &   & $t_5$ & 756 & 933 &   & Gumbel & 918 & 990 \\
    $X_2$ &   & Gumbel & 892 & 962 &   & $t_5$ & 686 & 911 &   & $t_5$ & 705 & 910 &   & Gumbel & 918 & 991 \\
    $X_3$ &   & Gumbel & 891 & 981 &   & $t_5$ & 711 & 915 &   & s.Gumbel & 901 & 980 &   & s.Gumbel & 902 & 982 \\
    $X_4$ &   & Gumbel & 900 & 984 &   & $t_5$ & 743 & 926 &   & s.Gumbel & 904 & 984 &   & s.Gumbel & 919 & 980 \\
    \bottomrule
    \end{tabular}%
\end{table}%

To  examine the reliability of using the heuristic  algorithm to select the true (simulated)   bivariate linking copulas, samples of size 500 were generated from various bi-factor and second-order copula models.
Table \ref{tab:model-selection} presents the number of times that the true (simulated) linking copulas were chosen over 1,000 simulation runs. It is revealed that the model selection algorithm performs extremely well for various bi-factor and second-order copulas models with different choices of linking copulas as the number of categories  $K$ increases.   For a small $K$ dependence in the tails cannot be easily quantified. Hence, for example, when the true copula is the $t_5$ which has the same upper and lower  tail dependence, the algorithm selected either $t_5$ or BVN  which  has zero lower and upper tail dependence, because both copulas provide reflection symmetric dependence.

To check whether the $\chi^2_{s-q}$  is a good approximation for the distribution of the $M_2$ statistic under the null hypothesis, samples of sizes 500 and 1000 were generated from various bi-factor second-order copula models. 
Table \ref{tab:M2} contains four common nominal levels of the $M_2$ statistic under the bi-factor and second-order copula models with different bivariate copulas. 
As can be seen in the table  the observed levels of $M_2$ are close to the nominal $\alpha$ levels and remain accurate even for extremely sparse tables ($d=16$ and $K=5$).

\begin{table}[!t]
  \centering
  \caption{\label{tab:M2}  Small sample of size $n =\{ 500,1000\}$ simulations (10$^3$ replications)  from  bi-factor and second-order copula models and  the empirical rejection levels at $\alpha = \{0.20, 0.10, 0.05, 0.01\}$, degrees of freedom (df), mean and variance. 
  }
    \begin{tabular}{lcccccccccc}
    \toprule
    \multicolumn{4}{l}{Bi-factor copula model} & \multicolumn{7}{c}{$M_2$} \\
\cmidrule{5-11}    Copula & $n$ & K &   & df & Mean & Variance &  $\alpha$=0.20 &  $\alpha$=0.10 &  $\alpha$=0.05 &  $\alpha$=0.01 \\
    \midrule
    BVN & 500 & 3 &   & 448 & 449.0 & 912.8 & 0.206 & 0.100 & 0.060 & 0.016 \\
      &   & 5 &   & 1888 & 1885.5 & 4858.3 & 0.210 & 0.117 & 0.065 & 0.024 \\
      & 1000 & 3 &   & 448 & 448.7 & 879.0 & 0.192 & 0.097 & 0.051 & 0.020 \\
      &   & 5 &   & 1888 & 1886.5 & 4332.5 & 0.202 & 0.108 & 0.064 & 0.015 \\
      &   &   &   &   &   &   &   &   &   &  \\
    Gumbel & 500 & 3 &   & 448 & 449.9 & 887.3 & 0.216 & 0.111 & 0.053 & 0.011 \\
      &   & 5 &   & 1888 & 1886.6 & 4709.7 & 0.225 & 0.126 & 0.070 & 0.015 \\
      & 1000 & 3 &   & 448 & 448.9 & 864.0 & 0.201 & 0.102 & 0.050 & 0.015 \\
      &   & 5 &   & 1888 & 1888.6 & 4332.1 & 0.226 & 0.107 & 0.069 & 0.014 \\
      &   &   &   &   &   &   &   &   &   &  \\
    $t_5$ & 500 & 3 &   & 448 & 448.7 & 907.3 & 0.202 & 0.088 & 0.048 & 0.018 \\
      &   & 5 &   & 1888 & 1886.6 & 4479.4 & 0.204 & 0.107 & 0.053 & 0.017 \\
      & 1000 & 3 &   & 448 & 448.6 & 834.9 & 0.184 & 0.090 & 0.050 & 0.014 \\
      &   & 5 &   & 1888 & 1890.3 & 4008.5 & 0.218 & 0.103 & 0.052 & 0.015 \\
    \bottomrule
  \multicolumn{4}{l}{Second-order  copula model} & \multicolumn{7}{c}{$M_2$} \\
\cmidrule{5-11}    Copula & $n$ & K &   & df & Mean & Variance &  $\alpha$=0.20 &  $\alpha$=0.10 &  $\alpha$=0.05 &  $\alpha$=0.01 \\
    \midrule
    BVN & 500 & 3 &   & 460 & 462.2 & 1001.2 & 0.220 & 0.113 & 0.055 & 0.016 \\
      &   & 5 &   & 1900 & 1903.5 & 3736.2 & 0.214 & 0.112 & 0.052 & 0.010 \\
      & 1000 & 3 &   & 460 & 461.3 & 1023.9 & 0.220 & 0.109 & 0.064 & 0.013 \\
      &   & 5 &   & 1900 & 1906.5 & 3918.2 & 0.230 & 0.130 & 0.068 & 0.012 \\
      &   &   &   &   &   &   &   &   &   &  \\
    Gumbel & 500 & 3 &   & 460 & 464.5 & 1011.3 & 0.233 & 0.117 & 0.073 & 0.024 \\
      &   & 5 &   & 1900 & 1909.2 & 5099.8 & 0.245 & 0.129 & 0.064 & 0.008 \\
      & 1000 & 3 &   & 460 & 461.9 & 871.2 & 0.203 & 0.106 & 0.049 & 0.009 \\
      &   & 5 &   & 1900 & 1908.5 & 3977.0 & 0.239 & 0.129 & 0.067 & 0.015 \\
      &   &   &   &   &   &   &   &   &   &  \\
    $t_5$ & 500 & 3 &   & 460 & 465.3 & 1362.4 & 0.247 & 0.145 & 0.091 & 0.039 \\
      &   & 5 &   & 1900 & 1904.7 & 3740.6 & 0.226 & 0.113 & 0.050 & 0.010 \\
      & 1000 & 3 &   & 460 & 461.8 & 900.1 & 0.214 & 0.108 & 0.055 & 0.010 \\
      &   & 5 &   & 1900 & 1908.1 & 3864.9 & 0.229 & 0.131 & 0.072 & 0.015 \\
    \bottomrule
    \end{tabular}%
\end{table}%

\baselineskip=25pt
	
\section{\label{sec-app}Application}
The Toronto Alexithymia Scale  is the most utilized measure of alexithymia in empirical research (\citealt{Bagby1994,Gignac-etal2007,Tuliao-etal2020}) 
and is composed of $d=20$ items that can be subdivided into $G=3$ non-overlapping groups:   $d_1=7$ items to assess difficulty identifying feelings (DIF),  $d_2=5$ items to assess  difficulty describing feelings (DDF) 
 and $d_3=8$ items to assess externally oriented thinking (EOT).  
We use a dataset of 1925 university students from the French-speaking region of Belgium \citep{Briganti&Linkowski2020}. Students were 17 to 25 years old and 58\% of them were female and 42\% were male. They were asked to respond to each item using one of $K=5$ categories: ``$1=$  completely disagree'', ``$2=$ disagree", ``$3=$ neutral, ``$4=$ agree", ``$5=$ completely agree". 
The dataset and  full description of the items can be  found  in the \proglang{R} package \pkg{ BGGM} \citep{BGGM_RPackage}.

For these items, a respondent might be thinking about the average ``sensation" of many past relevant events, leading to latent means. That is, based on the item descriptions, this seems more natural than a discretized maxima or minima. Since the sample is a mixture (male and female students) we can expect a priori that a bi-factor or second-order copula model with $t_\nu$ copulas might be plausible, as in this case the items can be considered as mixtures of discretized means.

\begin{table}[!h]
  \centering
  \small
    \caption{Average observed polychoric correlations and semi-correlations for all pairs  within each group and for all pairs of items for the Toronto Alexithymia Scale (TAS), along with the corresponding theoretical semi-correlations for BVN, $t_5$, Frank, Gumbel , and survival Gumbel (s.Gumbel) copulas. \label{tab:TASdisc}}
           \setlength{\tabcolsep}{7pt}  
    \begin{tabular}{lccccccccccccccc}
    \toprule
      & \multicolumn{3}{c}{All items} &   & \multicolumn{3}{c}{Items in group 1} &   & \multicolumn{3}{c}{Items in group 2} &   & \multicolumn{3}{c}{Items in group 3} \\
\cmidrule{2-4}\cmidrule{6-8}\cmidrule{10-12}\cmidrule{14-16}      & $\rho_N$ & $\rho_N^{-}$ & $\rho_N^{+}$ &   & $\rho_N$ & $\rho_N^{-}$ & $\rho_N^{+}$ &   & $\rho_N$ & $\rho_N^{-}$ & $\rho_N^{+}$ &   & $\rho_N$ & $\rho_N^{-}$ & $\rho_N^{+}$ \\
    \midrule
    Observed & 0.17 & 0.21 & 0.20 &   & 0.34 & 0.36 & 0.29 &   & 0.42 & 0.37 & 0.40 &   & 0.19 & 0.26 & 0.29 \\
    BVN & 0.17 & 0.07 & 0.07 &   & 0.34 & 0.16 & 0.16 &   & 0.42 & 0.21 & 0.21 &   & 0.19 & 0.08 & 0.08 \\
    $t_5$ & 0.17 & 0.23 & 0.23 &   & 0.34 & 0.31 & 0.31 &   & 0.42 & 0.35 & 0.35 &   & 0.19 & 0.24 & 0.24 \\
    Frank & 0.17 & 0.04 & 0.04 &   & 0.34 & 0.10 & 0.10 &   & 0.42 & 0.13 & 0.13 &   & 0.19 & 0.05 & 0.05 \\
    Gumbel & 0.17 & 0.05 & 0.22 &   & 0.34 & 0.11 & 0.37 &   & 0.42 & 0.14 & 0.43 &   & 0.19 & 0.05 & 0.24 \\
    s.Gumbel & 0.17 & 0.22 & 0.05 &   & 0.34 & 0.37 & 0.11 &   & 0.42 & 0.43 & 0.14 &   & 0.19 & 0.24 & 0.05 \\
    \bottomrule
    \end{tabular}%
\end{table}%

 In Table \ref{tab:TASdisc} we summarize the averages of  polychoric semi-correlations  for all pairs  within each group and for all pairs of items along with  the theoretical semi-correlations in (\ref{semicor}) under different choices of copulas.  For a BVN/$t_\nu$ copula the copula parameter is the sample polychoric correlation, while for a Gumbel/s.Gumbel copula the polychoric correlation  was converted to Kendall's tau with the relation in (\ref{tauBVN}) and then from Kendall's $\tau$  to Gumbel/s.Gumbel copula parameter via the functional inverse in (\ref{tauGumbel}). 
The summary of findings from the diagnostics  in the table show that 
\begin{itemize}
\item  for the first group of items there is more probability in the joint lower tail suggesting  s.Gumbel linking copulas
to join each item in this group with the DIF factor; 
\item  for the second group of items there is more probability in the joint lower and upper tail suggesting  $t_\nu$ linking copulas 
to join each item in this group with the DDF factor;
\item   for the third group of items there is more probability in the joint lower and upper tail suggesting  $t_\nu$ linking copulas 
to join each item in this group with the EOT factor;
\item for the  items overall there is more probability in the joint lower and upper tail suggesting  $t_\nu$ linking copulas 
to join each item or group specific factor (second-order model) with the common factor.
\end{itemize}
Hence, a bi-factor or second-order copula  model with the aforementioned  linking copulas might provide a better fit that the (Gaussian) models with BVN copulas.

Then, we fit the bi-factor and second-order models with 
the copulas selected  by the heuristic algorithm   in Section \ref{sec:selecalg}.  For a baseline comparison, we also fit their special cases;  these are the   one- and two-factor copula models where we have also selected the bivariate copulas using the heuristic algorithm proposed by \cite{Kadhem&Nikoloulopoulos2019}. To show the improvement of the copula models over their Gaussian analogues, we have also fitted all the classes of copula models with BVN copulas.    
The fitted models 
are compared via the
AIC, since the number of parameters is not the same between the models. In addition, we use the Vuong's test \citep{Vuong1989-Econometrica}
 to show if  (a) the best fitted model according to the AICs provides better fit than the other fitted models and (b) a model with the selected copulas provides better fit than the one with BVN copulas. 
The Vuong's test is the sample version of the difference in Kullback-Leibler divergence between two models and can be used to differentiate two  parametric models which could be non-nested. 
For the Vuong's test we provide the 95\% confidence interval of the Vuong's test statistic \citep[page 258]{Joe2014-CH}. If the interval does not contain  0, then the best fitted model according to the AICs is better if the interval is completely above 0. 
To assess the overall goodness-of-fit of the  bi-factor and second-order copula models, we  use the $M_2$ statistic \citep{Maydeu-Olivares&Joe2006}.

\begin{table}[!h]
\begin{minipage}{\textwidth}
  \centering
  \small
        \setlength{\tabcolsep}{2.5pt}  
 \caption{AICs, Vuong's 95\% CIs,
    and $M_2$ statistics for the 1-factor,  2-factor, bi-factor and second-order  copula models with BVN copulas and   selected copulas, along with the maximum deviations of observed and expected counts for all pairs within each group and for all pairs of items for the Toronto Alexithymia Scale. \label{tab:TASgf}}
    \begin{tabular}{lccccccccccc}
    \toprule
      & \multicolumn{2}{c}{1-factor} &   & \multicolumn{2}{c}{2-factor} &   & \multicolumn{2}{c}{Bi-factor} &   & \multicolumn{2}{c}{Second-order} \\
\cmidrule{2-3}\cmidrule{5-6}\cmidrule{8-9}\cmidrule{11-12}      & BVN & Selected &   & BVN & Selected &   & BVN & Selected &   & BVN & Selected \\
    \midrule
    AIC & 107135.8 & 105504.0 &   & 106189.5 & 103893.5 &   & 105507.7 & 103200.9 &   & 105878.6 & 104133.7 \\
    Vuong's 95\% CI\footnote{Selected factor copula model versus its Gaussian special case. } & \multicolumn{2}{c}{(0.35,0.50)} &   & \multicolumn{2}{c}{(0.53,0.69)} &   & \multicolumn{2}{c}{(0.51,0.69)} &   & \multicolumn{2}{c}{(0.38,0.52)} \\
    Vuong's 95\% CI\footnote{Selected Bi-factor copula model versus any other fitted model.} & (0.93,1.13) & (0.55,0.67) &   & (0.69,0.88) & (0.13,0.23) &   & \multicolumn{2}{c}{(0.51,0.69)} &   & (0.61,0.80) & (0.21,0.29) \\
    $M_2$ & 14723.8 & 9865.0 &   & 9195.7 & 7383.7 &   & 11664.7 & 6381.5 &   & 13547.1 & 7341.2 \\
    df & 3020 & 3020 &   & 3001 & 3000 &   & 3000 & 3000 &   & 3017 & 3017 \\
    $p$-value & $<0.001$ & $<0.001$ &   & $<0.001$ & $<0.001$ &   & $<0.001$ & $<0.001$ &   & $<0.001$ & $<0.001$ \\
    \midrule
    \multicolumn{4}{l}{Maximum discrepancy} &   &   &   &   &   &   &   &   \\
    \midrule
    Items in Group 1 & 71 & 63 &   & 71 & 60 &   & 69 & 55 &   & 70 & 61 \\
    Items in Group 2 & 112 & 98 &   & 113 & 83 &   & 77 & 48 &   & 84 & 55 \\
    Items in Group 3 & 87 & 74 &   & 81 & 52 &   & 80 & 45 &   & 82 & 53 \\
    All items & 112 & 98 &   & 113 & 83 &   & 80 & 55 &   & 84 & 61 \\
    \bottomrule
    \end{tabular}%
\begin{flushleft}
\end{flushleft}
\end{minipage}
\end{table}%

Table \ref{tab:TASgf}  gives the AICs, the 95\% CIs of Vuong's tests 
and the $M_2$ statistics for all the fitted models. 
The best fitted model,   based on AIC values, is the bi-factor copula model obtained from the selection algorithm. The best fitted bi-factor copula model results when we use s.Gumbel for the DIF factor,  $t_3$ for both the DDF and EOT factors and $t_2$ for the common factor (alexithymia).  
This is in line with the preliminary analyses based on the interpretations of items as mixtures of means and the diagnostics in Table \ref{tab:TASdisc}. It is revealed that the DIF items and DIF factor are discretized and latent minima, respectively, as the participants seem to reflect   that they ``disagree" or ``completely disagree"   having difficulty identifying feelings. 
From the Vuong's  95\% Cls and $M_2$ statistics  it is shown that  factor copula models provide  a big improvement over their Gaussian analogues and that the selected bi-factor copula model outperforms all the fitted models.

\begin{table}[!h]
  \centering
    \caption{Estimated copula parameters and their standard errors (SE) in Kendall’s $\tau$ scale for the Bi-factor copula models with BVN copulas and  selected copulas for the Toronto Alexithymia Scale. \label{tab:TASest}}
      \setlength{\tabcolsep}{5pt}  
    \begin{tabular}{lccccccclccclcc}
    \toprule
      &   & \multicolumn{5}{c}{Bi-factor copula model with BVN copulas} &   & \multicolumn{7}{c}{Bi-factor copula model with selected copulas} \\
\cmidrule{3-7}\cmidrule{9-15}      &   & \multicolumn{2}{c}{Common factor} &   & \multicolumn{2}{c}{Group-specific factors} &   & \multicolumn{3}{c}{Common factor} &   & \multicolumn{3}{c}{Group-specific factors} \\
    \midrule
    Items &   & Est & SE &   & Est & SE &   & Copulas & Est & SE &   & Copulas & Est & SE \\
    \midrule
    1 &   & 0.42 & 0.01 &   & 0.23 & 0.02 &   & $t_2$ & 0.49 & 0.02 &   & s.Gumbel & 0.09 & 0.03 \\
    3 &   & 0.14 & 0.02 &   & 0.24 & 0.02 &   & $t_2$ & 0.16 & 0.02 &   & s.Gumbel & 0.37 & 0.02 \\
    6 &   & 0.22 & 0.02 &   & 0.29 & 0.02 &   & $t_2$ & 0.29 & 0.02 &   & s.Gumbel & 0.23 & 0.02 \\
    7 &   & 0.11 & 0.02 &   & 0.31 & 0.02 &   & $t_2$ & 0.09 & 0.02 &   & s.Gumbel & 0.53 & 0.04 \\
    9 &   & 0.38 & 0.01 &   & 0.34 & 0.02 &   & $t_2$ & 0.47 & 0.02 &   & s.Gumbel & 0.24 & 0.02 \\
    13 &   & 0.36 & 0.01 &   & 0.46 & 0.02 &   & $t_2$ & 0.49 & 0.02 &   & s.Gumbel & 0.32 & 0.03 \\
    14 &   & 0.21 & 0.02 &   & 0.36 & 0.02 &   & $t_2$ & 0.30 & 0.02 &   & s.Gumbel & 0.27 & 0.03 \\
    \midrule
    2 &   & 0.71 & 0.02 &   & -0.24 & 0.10 &   & $t_2$ & 0.46 & 0.02 &   & $t_3$ & 0.53 & 0.02 \\
    4 &   & 0.55 & 0.01 &   & 0.02 & 0.04 &   & $t_2$ & 0.41 & 0.02 &   & $t_3$ & 0.58 & 0.03 \\
    11 &   & 0.35 & 0.01 &   & 0.13 & 0.03 &   & $t_2$ & 0.33 & 0.02 &   & $t_3$ & 0.20 & 0.03 \\
    12 &   & 0.34 & 0.02 &   & 0.29 & 0.04 &   & $t_2$ & 0.29 & 0.02 &   & $t_3$ & 0.23 & 0.03 \\
    17 &   & 0.31 & 0.02 &   & 0.38 & 0.06 &   & $t_2$ & 0.24 & 0.02 &   & $t_3$ & 0.25 & 0.03 \\
    \midrule
    5 &   & 0.06 & 0.02 &   & 0.33 & 0.02 &   & $t_2$ & 0.10 & 0.02 &   & $t_3$ & 0.34 & 0.02 \\
    8 &   & 0.11 & 0.02 &   & 0.30 & 0.02 &   & $t_2$ & 0.16 & 0.02 &   & $t_3$ & 0.33 & 0.02 \\
    10 &   & 0.12 & 0.02 &   & 0.27 & 0.02 &   & $t_2$ & 0.14 & 0.02 &   & $t_3$ & 0.30 & 0.02 \\
    15 &   & 0.15 & 0.02 &   & 0.19 & 0.02 &   & $t_2$ & 0.12 & 0.02 &   & $t_3$ & 0.19 & 0.02 \\
    16 &   & 0.03 & 0.02 &   & 0.23 & 0.02 &   & $t_2$ & 0.03 & 0.02 &   & $t_3$ & 0.24 & 0.02 \\
    18 &   & -0.02 & 0.02 &   & 0.28 & 0.02 &   & $t_2$ & 0.03 & 0.02 &   & $t_3$ & 0.29 & 0.02 \\
    19 &   & 0.07 & 0.02 &   & 0.40 & 0.02 &   & $t_2$ & 0.10 & 0.02 &   & $t_3$ & 0.43 & 0.02 \\
    20 &   & 0.06 & 0.02 &   & 0.27 & 0.02 &   & $t_2$ & 0.10 & 0.02 &   & $t_3$ & 0.26 & 0.02 \\
    \bottomrule
    \end{tabular}%
\end{table}%

Although the selected bi-factor copula model shows substantial improvement over the Gaussian bi-factor model or any other fitted model, it is not so clear from the goodness-of-fit $p$-values that the response patterns are satisfactorily explained by using the linking copulas selected by the heuristic algorithm. 
This is not surprising since one should expect discrepancies
between the postulated parametric model and the population probabilities, when the
sample size or dimension is sufficiently large \citep{Maydeu-Olivares&Joe2014}.
To further show that the fit has been improved we have calculated the maximum deviations of observed and model-based counts for each bivariate margin, that is,
$D_{j_1j_2}=n\max_{y_1,y_2}|p_{j_1,j_2,y_1,y_2}-\pi_{j_1,j_2,y_1,y_2}(\hat\thbf)|$. In Table \ref{tab:TASgf}  we   summarize the averages of these deviations for all pairs within each group and for all pairs of items.  Overall, the maximum discrepancies have been sufficiently reduced  in the selected bi-factor model.

Table \ref{tab:TASest}   gives the copula parameter estimates  in Kendall's $\tau$ scale and  their standard errors (SE) for the selected bi-factor copula model and the Gaussian bi-factor model as the benchmark model.  
The SEs of the estimated parameters are obtained by the inversion of the Hessian matrix at the second step of the IFM method.
These SEs are adequate to assess the flatness of the log-likelihood.
Proper SEs that account for the estimation of cutpoints can be obtained by
jackknifing the two-stage estimation procedure.
The loading parameters ($\hat\tau$'s converted to BVN copula parameters  via the functional inverse in (\ref{tauBVN}) and then to loadings using the relations in Section 2.3) show that 
 the common alexithymia factor is mostly loaded  on DIF and DDF items,  suggesting that items in the domains DIF and DDF are good indicators for alexithymia. 
 The items in the EOT although they loaded on the EOT latent factor, they had poor loadings in the common alexithymia factor.

\section{\label{sec-disc}Discussion}
For item response data that can be split into non-overlapping groups, we have proposed bi-factor and second-order copula models  where we replace BVN distributions, between observed and latent variables, with bivariate copulas. Our  copula constructions include the Gaussian bi-factor and second-order models as special cases and can provide a substantial improvement over the Gaussian models  based on AIC, Vuong's and goodness-of-fit statistics. Hence, superior statistical inference for the loading parameters of interest can be achieved. The improvement relies on the fact that when we use appropriate bivariate copulas other than BVN copulas in the construction, there is an interpretation of latent variables that can be maxima/minima or mixture of means instead of means.

Our constructions have a latent structure that is not additive as in (\ref{eq:bifactorBVN}) and (\ref{2nd-order-gaussian}) if other than BVN copulas are called and the bi-factor copula (dependence) parameters are interpretable as dependence of an observed variable with the common factor, or conditional dependence of an observed variable with  the group-specific latent variable given the common factor.

We have proposed a fast and efficient likelihood estimation technique based on  Gauss-Legendre quadrature points.   The joint pmfs in (\ref{bifactor-pmf}) and (\ref{2ndorder-pmf}) reduce to  one-dimensional integrals of a function which in turn is a product of $G$ one-dimensional integrals.  Hence, 
the evaluation of the  the joint likelihood  requires only low-dimensional integration regardless of the dimension of the latent variables. 

Building on the  models proposed in this paper, there are several extensions that can be implemented. The adoption of the  structure of the Gaussian  tri-factor and the third-order  models (e.g., \citealt{Rijmen-etal2014}), to account for any additional layer of dependence,   is feasible using the notion of
 truncated 
vine copulas that involve both observed and latent variables.

\section*{Software}
\proglang{R} functions for estimation, model selection and goodness-of-fit of the bi-factor and second-order copula models will be part of the next major release of the \proglang{R} package \pkg{FactorCopula} \citep{Kadhem&Nikoloulopoulos-2020-package}.

\section*{Acknowledgements}
 The simulations presented in this paper were carried out on the High Performance Computing Cluster supported by the Research and Specialist Computing Support service at the University of East Anglia.

\section*{Appendix}
We provide the form  of   the 
 derivatives of the univariate and bivariate marginal probabilities with respect to the estimated model parameters in the Appendix Tables 1--5. 

\bigskip

\captionsetup[table]{name=Appendix Table}
\setcounter{table}{0}

 \begin{table}[!h]
  \centering
  \small
	\caption{ \label{tab:M2derivativesUNIV} Derivatives of the univariate probability $\pi_{jg,y}=\Phi(\alpha_{jg,y+1}) - \Phi(\alpha_{jg,y})$ with respect to the cutpoint $\alpha_{jg,k}$ for $g=1\ldots,G, \;  j = 1,\ldots, d_g, \;  y = 1, \ldots, K - 1, \mbox{   and  }  k = 1,\ldots, K - 1$. }
	    \setlength{\tabcolsep}{95pt}  
    \begin{tabular}{ll}
    \toprule
    $\partial \pi_{jg,y}/\partial \alpha_{jg,k} $ & If  \\
    \midrule\\
	    $\phi(\alpha_{jg,y+1})$ & $ k=y+1$ \\\\
	    	$- \phi(\alpha_{jg,y})$ & $ k=y$ \\\\
    \bottomrule
    \end{tabular}
    \end{table}

\begin{sidewaystable}[htbp]
  \centering
  \small
	\caption{\label{tab:M2derivativesBIF1} Derivatives of the bivariate probability $\pi_{j_1j_2g,y_1,y_2}=\Pr(Y_{j_1g}=y_1,Y_{j_2g}=y_2)$ with respect to the cutpoint $\alpha_{jg,k}$, the copula parameter $\theta_{jg}$ for the common factor $X_0$, and the copula parameter $\delta_{jg}$ for the group-specific factor $X_g$ for the bi-factor copula model for $g=1\ldots,G, \;  j, j_1, j_2 = 1,\ldots, d_g, \; y, y_1, y_2 = 1, \ldots, K - 1, \mbox{   and  }  k = 1,\ldots, K - 1$.  Note that
$f_{Y_{jg}|X_{g};X_0}(y_{jg}|x_g;x_0) = \Big( C_{Y_{jg}|X_g;X_0}\bigl(   C_{Y_{jg}|X_0}(a_{jg,y+1}|x_0; \theta_{jg})   | x_g; \delta_{jg} \bigr)  -C_{Y_{jg}|X_g;X_0}\bigr(    C_{Y_{jg}|X_0}(a_{jg,y}|x_0; \theta_{jg})     | x_g; \delta_{jg}\bigr) \Big)$ where $a_{jg,k} =\Phi(\alpha_{jg,k})$, \quad  
$c_{X_0Y_{jg}}(x_0, a) = \partial^2 C_{X_0 Y_{jg}}(x_0,a)/\partial x_0 \partial a  $,  \quad  
$\dot{C}_{jg|X_{0}}(\cdot;\theta_{jg}) = \partial C_{jg|X_{0}}(\cdot; \theta_{jg})/\partial \theta_{jg}$, \quad  
$\dot{C}_{Y_{jg}|X_{g}; X_0}(\cdot;\delta_{jg}) = \partial C_{Y_{jg}|X_{g};X_0}(\cdot; \delta_{jg})/\partial \delta_{jg}$, \quad  
$\dot{f}_{Y_{jg}|X_{jg};X_0}(y_{jg}|x_g;x_0) = \partial f_{Y_{jg}|X_{jg};X_0}(y_{jg}|x_g;x_0)/\partial \delta_{jg} =   \dot{C}_{Y_{jg}|X_g;X_0}\bigl(   C_{Y_{jg}|X_0}(a_{jg,y+1}|x_0)   | x_g\bigr)  - \dot{C}_{Y_{jg}|X_g;X_0}\bigr(    C_{Y_{jg}|X_0}(a_{jg,y}|x_0)     | x_g\bigr)$, \quad 
$\bar{f}_{Y_{jg}|X_{jg};X_0}(y_{jg}|x_g;x_0) = \partial f_{Y_{jg}|X_{jg};X_0}(y_{jg}|x_g;x_0)/\partial \theta_{jg} =   c_{X_gY_{jg}}\bigl(x_g, C_{Y_{jg}|X_0}(a_{jg,y+1}| x_0)\bigr) \dot{C}_{Y_{jg}|X_0}(a_{jg,y+1}| x_0) - c_{X_gY_{jg}}\bigr(x_g,  C_{Y_{jg}|X_0}(a_{jg,y}|x_0)  \bigr) \dot{C}_{Y_{jg}|X_0}(a_{jg,y}|x_0). $
	}
    \setlength{\tabcolsep}{40pt}  
    \begin{tabular}{ll}
    \toprule
    $\partial \pi_{j_1j_2g,y_1,y_2}/\partial \alpha_{jg,k} $& If\\
    \midrule
\\
$\phi(\alpha_{j_1g,y_1+1})    
\int_0^1   \int_0^1  \;\;
f_{Y_{j_2g}|X_{g};X_0}(y_{j_2g}|x_g;x_0)  \;\;
c_{X_gY_{j_1g}}\bigl(x_g, C_{Y_{j_1g}|X_0}(a_{j_1g,y_1+1}| x_0)\bigr)\;\;
 c_{X_0Y_{j_1g}}(x_0,a_{j_1g,y_1+1})\;\;
  dx_g    dx_0$  & $ j=j_1, k=y_1+1$\\\\
$-\phi(\alpha_{j_1g,y_1})    \;
\int_0^1   \int_0^1  \;\;
f_{Y_{j_2g}|X_{g};X_0}(y_{j_2g}|x_g;x_0)   \;\;
c_{X_gY_{j_1g}}\bigl(x_g, C_{Y_{j_1g}|X_0}(a_{j_1g,y_1}| x_0)\bigr)\;\;
 c_{X_0Y_{j_1g}}(x_0,a_{j_1g,y_1})\;\;
  dx_g    dx_0$  & $ j=j_1, k=y_1$\\\\
$\phi(\alpha_{j_2g,y_2+1})    
\int_0^1   \int_0^1 \;\;
f_{Y_{j_1g}|X_{g};X_0}(y_{j_1g}|x_g;x_0)  \;\;
c_{X_gY_{j_2g}}\bigl(x_g, C_{Y_{j_2g}|X_0}(a_{j_2g,y_2+1}| x_0)\bigr)\;\;
 c_{X_0Y_{j_2g}}(x_0,a_{j_2g,y_2+1})\;\;
  dx_g    dx_0$  & $ j=j_2, k=y_2+1$\\\\
$-\phi(\alpha_{j_2g,y_2})    \;
\int_0^1   \int_0^1 \;\;
f_{Y_{j_1g}|X_{g};X_0}(y_{j_1g}|x_g;x_0)  \;\;
c_{X_gY_{j_2g}}\bigl(x_g, C_{Y_{j_2g}|X_0}(a_{j_2g,y_2}| x_0)\bigr)\;\;
 c_{X_0Y_{j_2g}}(x_0,a_{j_2g,y_2})\;\;
  dx_g    dx_0$  & $ j=j_2, k=y_2$\\\\
  
      \toprule
    $\partial \pi_{j_1j_2g,y_1,y_2}/\partial \theta_{jg} $& If\\
    \midrule
\\
 $ \int_0^1   \int_0^1  \;\;
f_{Y_{j_2g}|X_{g};X_0}(y_{j_2g}|x_g;x_0)\;\;
\bar{f}_{Y_{j_1g}|X_{j_1g};X_0}(y_{j_1g}|x_g;x_0)\;\;
 dx_g   dx_0 $
& $ j =   j_1 $\\\\
 $ \int_0^1   \int_0^1  \;\;
f_{Y_{j_1g}|X_{g};X_0}(y_{j_1g}|x_g;x_0)\;\;
\bar{f}_{Y_{j_2g}|X_{j_2g};X_0}(y_{j_2g}|x_g;x_0)\;\;
 dx_g   dx_0 $
& $ j =   j_2 $\\\\
      \toprule
    $\partial \pi_{j_1j_2g,y_1,y_2}/\partial \delta_{jg} $& If\\
    \midrule
\\
 $ \int_0^1   \int_0^1  \;\;
f_{Y_{j_2g}|X_{g};X_0}(y_{j_2g}|x_g;x_0)\;\;
\dot{f}_{Y_{j_1g}|X_{j_1g};X_0}(y_{j_1g}|x_g;x_0)\;\;
 dx_g   dx_0 $
& $ j =  j_1 $\\\\

 $ \int_0^1   \int_0^1  \;\;
f_{Y_{j_1g}|X_{g};X_0}(y_{j_1g}|x_g;x_0)\;\;
\dot{f}_{Y_{j_2g}|X_{j_2g};X_0}(y_{j_2g}|x_g;x_0)\;\;
 dx_g   dx_0 $
& $ j =   j_2 $\\\\
  \bottomrule
\end{tabular}
\end{sidewaystable}

\begin{sidewaystable}
\centering
\small
	\caption{\label{tab:M2derivativesBIF2} Derivatives of the bivariate probability $\pi_{j_1g_1j_2g_2,y_1,y_2}=\Pr(Y_{j_1g_1}=y_1,Y_{j_2g_2}=y_2)$ with respect to the cutpoint $\alpha_{jg,k}$, the copula parameter $\theta_{jg}$ for the common factor $X_0$, and the copula parameter $\delta_{jg}$ for the group-specific factor $X_g$ for the bi-factor copula model for $g=1\ldots,G, \;  j, j_1, j_2 = 1,\ldots, d_g, \; y, y_1, y_2 = 1, \ldots, K - 1, \mbox{   and  }  k = 1,\ldots, K - 1$.  Note that 
$f_{Y_{jg}|X_{g};X_0}(y_{jg}|x_g;x_0) = \Big( C_{Y_{jg}|X_g;X_0}\bigl(   C_{Y_{jg}|X_0}(a_{jg,y+1}|x_0; \theta_{jg})   | x_g; \delta_{jg} \bigr)  -C_{Y_{jg}|X_g;X_0}\bigr(    C_{Y_{jg}|X_0}(a_{jg,y}|x_0; \theta_{jg})     | x_g; \delta_{jg}\bigr) \Big)$ where $a_{jg,k} =\Phi(\alpha_{jg,k})$, \quad  
$c_{X_0Y_{jg}}(x_0, a) = \partial^2 C_{X_0 Y_{jg}}(x_0,a)/\partial x_0 \partial a  $,  \quad  
$\dot{C}_{jg|X_{0}}(\cdot;\theta_{jg}) = \partial C_{jg|X_{0}}(\cdot; \theta_{jg})/\partial \theta_{jg}$, \quad  
$\dot{C}_{Y_{jg}|X_{g}; X_0}(\cdot;\delta_{jg}) = \partial C_{Y_{jg}|X_{g};X_0}(\cdot; \delta_{jg})/\partial \delta_{jg}$, \quad  
$\dot{f}_{Y_{jg}|X_{jg};X_0}(y_{jg}|x_g;x_0) = \partial f_{Y_{jg}|X_{jg};X_0}(y_{jg}|x_g;x_0)/\partial \delta_{jg} =   \dot{C}_{Y_{jg}|X_g;X_0}\bigl(   C_{Y_{jg}|X_0}(a_{jg,y+1}|x_0)   | x_g\bigr)  - \dot{C}_{Y_{jg}|X_g;X_0}\bigr(    C_{Y_{jg}|X_0}(a_{jg,y}|x_0)     | x_g\bigr)$, \quad 
$\bar{f}_{Y_{jg}|X_{jg};X_0}(y_{jg}|x_g;x_0) = \partial f_{Y_{jg}|X_{jg};X_0}(y_{jg}|x_g;x_0)/\partial \theta_{jg} =   c_{X_gY_{jg}}\bigl(x_g, C_{Y_{jg}|X_0}(a_{jg,y+1}| x_0)\bigr) \dot{C}_{Y_{jg}|X_0}(a_{jg,y+1}| x_0) - c_{X_gY_{jg}}\bigr(x_g,  C_{Y_{jg}|X_0}(a_{jg,y}|x_0)  \bigr) \dot{C}_{Y_{jg}|X_0}(a_{jg,y}|x_0). $
	}
	
   \setlength{\tabcolsep}{16pt}  
\begin{tabular}{ll}
\toprule
$\partial \pi_{j_1g_1j_2g_2,y_1,y_2}/\partial \alpha_{jg,k} $ & If\\
\midrule
\\
$\phi(\alpha_{j_1g_1,y_1+1})    
\int_0^1   \int_0^1 \;\;
f_{Y_{j_2g_2}|X_{g_2};X_0}(y_{j_2g_2}|x_{g_2};x_0) \;\;
  dx_{g_2} 
  \int_0^1 \;\;
c_{X_{g_1}Y_{j_1g_1}}\bigl(x_{g_1}, C_{Y_{j_1g_1}|X_0}(a_{j_1g_1,y_1+1}| x_0)\bigr) \;\;
 c_{X_0Y_{j_1g_1}}(x_0,a_{j_1g_1,y_1+1}) \;\;
  dx_{g_1}     dx_0$  & $ j=j_1, g=g_1, k=y_1+1$\\\\
$-\phi(\alpha_{j_1g_1,y_1})    
\int_0^1   \int_0^1  \;\;
f_{Y_{j_2g_2}|X_{g_2};X_0}(y_{j_2g_2}|x_{g_2};x_0)  \;\;
  dx_{g_2} 
\int_0^1  \;\;
c_{X_{g_1}Y_{j_1g_1}}\bigl(x_{g_1}, C_{Y_{j_1g_1}|X_0}(a_{j_1g_1,y_1}| x_0)\bigr) \;\;
 c_{X_0Y_{j_1g_1}}(x_0,a_{j_1g_1,y_1}) \;\;
  dx_{g_1}    dx_0$  & $ j=j_1, g=g_1, k=y_1$\\\\
  $\phi(\alpha_{j_2g_2,y_2+1})    
\int_0^1   \int_0^1  \;\;
f_{Y_{j_1g_1}|X_{g_1};X_0}(y_{j_1g_1}|x_{g_1};x_0)   \;\;
dx_{g_1} 
\int_0^1  \;\;
c_{X_{g_2}Y_{j_2g_2}}\bigl(x_{g_2}, C_{Y_{j_2g_2}|X_0}(a_{j_2g_2,y_2+1}| x_0)\bigr) \;\;
 c_{X_0Y_{j_2g_2}}(x_0,a_{j_2g_2,y_2+1}) \;\;
  dx_{g_2}    dx_0$  & $ j=j_2, g=g_2, k=y_2+1$\\\\
$-\phi(\alpha_{j_2g_2,y_2})    
\int_0^1   \int_0^1  \;\;
f_{Y_{j_1g_1}|X_{g_1};X_0}(y_{j_1g_1}|x_{g_1};x_0)   \;\;
dx_{g_1}  
 \int_0^1   \;\;
c_{X_{g_2}Y_{j_2g_2}}\bigl(x_{g_2}, C_{Y_{j_2g_2}|X_0}(a_{j_2g_2,y_2}| x_0)\bigr) \;\;
 c_{X_0Y_{j_2g_2}}(x_0,a_{j_2g_2,y_2}) \;\;
  dx_{g_2}    dx_0$  &   $ j=j_2, g=g_2, k=y_2$\\\\
  
\toprule
$\partial \pi_{j_1g_1j_2g_2,y_1,y_2}/\partial \theta_{jg} $ & If\\
\midrule
\\
 $ \int_0^1   \int_0^1   \;\;
f_{Y_{j_2g_2}|X_{g_2};X_0}(y_{j_2g_2}|x_{g_2};x_0) \;\;
 dx_{g_2} 
 \int_0^1   \;\;
 \bar{f}_{Y_{j_1g_1}|X_{j_1g_1};X_0}(y_{j_1g_1}|x_{g_1};x_0)\;\;
 dx_{g_1}    dx_0 $
& $ j =j_1, g=g_1 $		\\\\ 
 $ \int_0^1   \int_0^1  \;\; 
f_{Y_{j_1g_1}|X_{g_1};X_0}(y_{j_1g_1}|x_{g_1};x_0)  \;\;
dx_{g_1} 
\int_0^1   \;\;
\bar{f}_{Y_{j_2g_2}|X_{j_2g_2};X_0}(y_{j_2g_2}|x_{g_2};x_0) \;\;
 dx_{g_2}   dx_0 $
& $ j = j_2 , g=g_2$\\\\

\toprule
$\partial \pi_{j_1g_1j_2g_2,y_1,y_2}/\partial \delta_{jg} $ & If\\
\midrule
\\
 $ \int_0^1   \int_0^1   \;\;
f_{Y_{j_2g_2}|X_{g_2};X_0}(y_{j_2g_2}|x_{g_2};x_0)   \;\;
dx_{g_2}
\int_0^1  \;\;
\dot{f}_{Y_{j_1g_1}|X_{j_1g_1};X_0}(y_{j_1g_1}|x_{g_1};x_0) \;\;
 dx_{g_1}  dx_0 $
& $ j =   j_1, g=g_1 $\\\\
 $ \int_0^1    \;\;
\int_0^1  f_{Y_{j_1g_1}|X_{g_1};X_0}(y_{j_1g_1}|x_{g_1};x_0)  \;\;
dx_{g_1} 
\int_0^1   \;\;
\dot{f}_{Y_{j_2g_2}|X_{j_2g_2};X_0}(y_{j_2g_2}|x_{g_2};x_0) \;\;
 dx_{g_2}   
 dx_0 $
& $ j =   j_2 , g=g_2$\\\\
  \bottomrule
\end{tabular}
\end{sidewaystable}

\begin{sidewaystable}[htbp]
  \centering
  \small
\caption{\label{tab:M2derivativesScndOrdr1} Derivatives of the bivariate probabilities $\pi_{j_1j_2g,y_1,y_2}=\Pr(Y_{j_1g}=y_1,Y_{j_2g}=y_2)$  with respect to the cutpoint $\alpha_{jg,k}$, the copula parameter $\theta_{jg}$ for the first-order factor $X_g$, and the copula parameter $\delta_{g}$ for the the second-order factor $X_0$ for the second-order  copula model for $g=1\ldots,G, \;  j, j_1, j_2 = 1,\ldots, d_g, \; y, y_1, y_2 = 1, \ldots, K - 1, \mbox{   and  }  k = 1,\ldots, K - 1$.  Note that 
$
f_{Y_{jg}|X_g}(y_{jg}|x_g) =  C_{Y_{jg}|X_g}\bigl(a_{jg,y+1} | x_g; \theta_{jg}\bigr) - C_{Y_{jg}|X_g}\bigl(a_{jg,y} | x_g; \theta_{jg} \bigr),
$\quad
$c_{X_gY_{jg}}(x_g, a) = \partial^2 C_{X_g Y_{jg}}(x_g,a)/\partial x_g \partial a,$\quad  
$\dot{C}_{Y_{jg}|X_{g}}(\cdot;\theta_{jg}) = \partial C_{Y_{jg}|X_{g}}(\cdot; \theta_{jg})/\partial \theta_{jg}$, \quad  
$\dot{f}_{Y_{jg}|X_{jg}}(y_{jg}|x_g) = \partial f_{Y_{jg}|X_{jg}}(y_{jg}|x_g)/\partial \theta_{jg} =   \dot{C}_{Y_{jg}|X_g}\bigl(  a_{jg,y+1}  | x_g\bigr)  - \dot{C}_{Y_{jg}|X_g}\bigl(  a_{jg,y}  | x_g\bigr),$ \quad 
$\dot{c}_{X_gX_0}(x_g,x_0; \delta_g) = \partial c_{X_gX_0}(x_g,x_0; \delta_g)/\partial \delta_{g}$.
}
    \setlength{\tabcolsep}{72pt}  
    \begin{tabular}{ll}
    \toprule
       $\partial \pi_{j_1j_2g,y_1,y_2}/\partial \alpha_{jg,k} $& If\\
    \midrule
    \\
$\phi(\alpha_{j_1g,y_1+1})  \int_0^1   \int_0^1  \;\;
f_{Y_{j_2g}|X_g}(y_{j_2g}|x_g)   \;\;
c_{X_gY_{j_1g}}(x_g, a_{j_1g,y_1+1})   \;\;
 c_{X_gX_0}(x_g,x_0)  \;\;
  dx_g    dx_0 $ &  $j=j_1, k=y_1+1$\\\\
$-\phi(\alpha_{j_1g,y_1})  \; \int_0^1   \int_0^1  \;\;
f_{Y_{j_2g}|X_g}(y_{j_2g}|x_g)   \;\;
c_{X_gY_{j_1g}}(x_g, a_{j_1g,y_1})  \;\;
 c_{X_gX_0}(x_g,x_0)  \;\;
  dx_g    dx_0 $ &  $j=j_1, k=y_1$\\\\
$\phi(\alpha_{j_2g,y_2+1})  \int_0^1   \int_0^1  \;\;
f_{Y_{j_1g}|X_g}(y_{j_1g}|x_g)   \;\;
c_{X_gY_{j_2g}}(x_g, a_{j_2g,y_2+1}) \;\;
 c_{X_gX_0}(x_g,x_0) \;\;
  dx_g    dx_0 $ &  $j=j_2, k=y_2+1$\\\\
$-\phi(\alpha_{j_2g,y_2})  \;	\int_0^1   \int_0^1 \;\;
f_{Y_{j_1g}|X_g}(y_{j_1g}|x_g)  \;\;
c_{X_gY_{j_2g}}(x_g, a_{j_2g,y_2})  \;\;
 c_{X_gX_0}(x_g,x_0) \;\;
  dx_g    dx_0 $ &  $j=j_2, k=y_2$\\\\
      \toprule
       $\partial \pi_{j_1j_2g,y_1,y_2}/\partial \theta_{jg} $& If\\
    \midrule
    \\
 $  \int_0^1   \int_0^1  \;\;
f_{Y_{j_2g}|X_g}(y_{j_2g}|x_g)  \;\;
\dot{f}_{Y_{j_1g}|X_g}(y_{j_1g}|x_g)  \;\;
 c_{X_gX_0}(x_g,x_0) \;\;
    dx_g   dx_0 $ & $ j =   j_1$\\\\
 $  \int_0^1   \int_0^1  \;\;
f_{Y_{j_1g}|X_g}(y_{j_1g}|x_g)  \;\;
\dot{f}_{Y_{j_2g}|X_g}(y_{j_2g}|x_g)  \;\;
 c_{X_gX_0}(x_g,x_0) \;\;
    dx_g   dx_0 $ & $ j =   j_2 $\\\\
      \toprule
       $\partial \pi_{j_1j_2g,y_1,y_2}/\partial \delta_{g} $& \\
    \midrule
    \\
$ \int_0^1   \int_0^1     \;\;
f_{Y_{j_1g}|X_g}(y_{j_1g}|x_g)  \;\;
f_{Y_{j_2g}|X_g}(y_{j_2g}|x_g) \;\;
\dot{c}_{X_gX_0}(x_g,x_0) \;\;
 dx_g   dx_0$ 
 &  \\\\
 \bottomrule
\end{tabular}
\end{sidewaystable} 
 
 \begin{sidewaystable} 
   \centering
   \small
 \caption{\label{tab:M2derivativesScndOrdr2} Derivatives of the bivariate probability $\pi_{j_1g_1j_2g_2,y_1,y_2}=\Pr(Y_{j_1g_1}=y_1,Y_{j_2g_2}=y_2)$  with respect to the cutpoint $\alpha_{jg,k}$, the copula parameter $\theta_{jg}$ for the first-order factor $X_g$, and the copula parameter $\delta_{g}$ for the second-order factor $X_0$ for the second-order copula model for $g=1\ldots,G, \;  j, j_1, j_2 = 1,\ldots, d_g, \; y, y_1, y_2 = 1, \ldots, K - 1, \mbox{   and  }  k = 1,\ldots, K - 1$.  Note that 
$
f_{Y_{jg}|X_g}(y_{jg}|x_g) =  C_{Y_{jg}|X_g}\bigl(a_{jg,y+1} | x_g; \theta_{jg}\bigr) - C_{Y_{jg}|X_g}\bigl(a_{jg,y} | x_g; \theta_{jg} \bigr),
$\quad
$c_{X_gY_{jg}}(x_g, a) = \partial^2 C_{X_g Y_{jg}}(x_g,a)/\partial x_g \partial a,$\quad  
$\dot{C}_{Y_{jg}|X_{g}}(\cdot;\theta_{jg}) = \partial C_{Y_{jg}|X_{g}}(\cdot; \theta_{jg})/\partial \theta_{jg}$, \quad  
$\dot{f}_{Y_{jg}|X_{jg}}(y_{jg}|x_g) = \partial f_{Y_{jg}|X_{jg}}(y_{jg}|x_g)/\partial \theta_{jg} =   \dot{C}_{Y_{jg}|X_g}\bigl(  a_{jg,y+1}  | x_g\bigr)  - \dot{C}_{Y_{jg}|X_g}\bigl(  a_{jg,y}  | x_g\bigr),$ \quad 
$\dot{c}_{X_gX_0}(x_g,x_0; \delta_g) = \partial c_{X_gX_0}(x_g,x_0; \delta_g)/\partial \delta_{g}$.
}
    \setlength{\tabcolsep}{23pt}  
 \begin{tabular}{ll}
\toprule
$\partial \pi_{j_1g_1j_2g_2,y_1,y_2}/\partial \alpha_{jg,k} $ & If\\
\midrule
\\
$\phi(\alpha_{j_1g_1,y_1+1})  \int_0^1   
\int_0^1 \;\;
f_{Y_{j_2g_2}|X_{g_2}}(y_{j_2g_2}|x_{g_2})  \;\;
 c_{X_{g_2}X_0}(x_{g_2},x_0) \;\;
dx_{g_2}    
\int_0^1 \;\;
c_{X_{g_1}Y_{j_1{g_1}}}(x_{g_1}, a_{j_1g_1,y_1+1})  \;\;
 c_{X_{g_1}X_0}(x_{g_1},x_0) \;\;
dx_{g_1}   dx_0 $ &  $j=j_1, g=g_1, k=y_1+1$\\\\
$-\phi(\alpha_{j_1g_1,y_1})  \; \int_0^1   
\int_0^1 \;\;
f_{Y_{j_2g_2}|X_{g_2}}(y_{j_2g_2}|x_{g_2})  \;\;
 c_{X_{g_2}X_0}(x_{g_2},x_0) \;\;
dx_{g_2}    
\int_0^1 \;\;
c_{X_{g_1}Y_{j_1g_1}}(x_{g_1}, a_{j_1g_1,y_1})  \;\;
 c_{X_{g_1}X_0}(x_{g_1},x_0) \;\;
dx_{g_1}    
 dx_0 $ &  $j=j_1, g=g_1, k=y_1$\\\\
$\phi(\alpha_{j_2g_2,y_2+1})  \int_0^1   
\int_0^1 \;\;
f_{Y_{j_1g_1}|X_{g_1}}(y_{j_1g_1}|x_{g_1})  \;\;
 c_{X_{g_1}X_0}(x_{g_1},x_0) \;\;
dx_{g_1}    
\int_0^1 \;\;
c_{X_{g_2}Y_{j_2g_2}}(x_{g_2}, a_{j_2g_2,y_2+1})  \;\;
 c_{X_{g_2}X_0}(x_{g_2},x_0)\;\;
dx_{g_2}    
 dx_0 $ &  $j=j_2, g=g_2, k=y_2+1$\\\\
$-\phi(\alpha_{j_2g_2,y_2})  \; \int_0^1   
\int_0^1 \;\;
f_{Y_{j_1g_1}|X_{g_1}}(y_{j_1g_1}|x_{g_1})   \;\;
 c_{X_{g_1}X_0}(x_{g_1},x_0) \;\;
dx_{g_1}    
\int_0^1 \;\;
c_{X_{g_2}Y_{j_2g_2}}(x_{g_2}, a_{j_2g_2,y_2})   \;\;
 c_{X_{g_2}X_0}(x_{g_2},x_0) \;\;
dx_{g_2}    
 dx_0 $ &  $j=j_2, g=g_2, k=y_2$\\\\
\toprule
$\partial \pi_{j_1g_1j_2g_2,y_1,y_2}/\partial \theta_{jg} $ & If\\
\midrule
\\
 $  \int_0^1   \int_0^1  \;\;
f_{Y_{j_2g_2}|X_{g_2}}(y_{j_2g_2}|x_{g_2})    \;\;
 c_{X_{g_2}X_0}(x_{g_2},x_0) \;\;
    dx_{g_2}   
 \int_0^1  \;\;
\dot{f}_{Y_{j_1g_1}|X_{g_1}}(y_{j_1g_1}|x_{g_1})    \;\;
 c_{X_{g_1}X_0}(x_{g_1},x_0) \;\;
    dx_{g_1}   
    dx_0 $ & $ j=j_1, g=g_1 $\\\\
 $  \int_0^1    \int_0^1  \;\;
f_{Y_{j_1g_1}|X_{g_1}}(y_{j_1g_1}|x_{g_1})   \;\;
 c_{X_{g_1}X_0}(x_{g_1},x_0) \;\;
    dx_{g_1}   
 \int_0^1  \;\;
\dot{f}_{Y_{j_2g_2}|X_{g_2}}(y_{j_2g_2}|x_{g_2})   \;\;
 c_{X_{g_2}X_0}(x_{g_2},x_0) \;\;
    dx_{g_2}
    dx_0 $ & $ j=j_2, g=g_2 $\\\\
    
\toprule
$\partial \pi_{j_1g_1j_2g_2,y_1,y_2}/\partial \delta_{g} $ & If\\
\midrule
\\
$ \int_0^1  \int_0^1   \;\;
f_{Y_{j_2g_2}|X_{g_2}}(y_{j_2g_2}|x_{g_2})    \;\;
c_{X_{g_2}X_0}(x_{g_2},x_0) \;\;
 dx_{g_2}   
  \int_0^1     \;\;
f_{Y_{j_1g_1}|X_{g_1}}(y_{j_1g_1}|x_{g_1})    \;\;
\dot{c}_{X_{g_1}X_0}(x_{g_1},x_0) \;\;
 dx_{g_1}    dx_0$  &  $g =   g_1 $\\\\
$ \int_0^1   \int_0^1     \;\;
f_{Y_{j_1g_1}|X_{g_1}}(y_{j_1g_1}|x_{g_1})    \;\;
c_{X_{g_1}X_0}(x_{g_1},x_0) \;\;
 dx_{g_1}   
 \int_0^1   \;\;
f_{Y_{j_2g_2}|X_{g_2}}(y_{j_2g_2}|x_{g_2}) \;\;
\dot{c}_{X_{g_2}X_0}(x_{g_2},x_0) \;\;
 dx_{g_2}   dx_0$  &  $g =   g_2 $\\\\
  \bottomrule
\end{tabular}
\end{sidewaystable}


\end{document}